\documentclass[fleqn,usenatbib]{rasti}

\usepackage{newtxtext,newtxmath}
\usepackage[T1]{fontenc}
\DeclareRobustCommand{\VAN}[3]{#2}
\let\VANthebibliography\thebibliography
\def\thebibliography{\DeclareRobustCommand{\VAN}[3]{##3}\VANthebibliography}

\usepackage{float}
\restylefloat{table}
\restylefloat{figure}
\usepackage{graphicx}	
\usepackage{amsmath}	
\usepackage{bm}

\usepackage{caption}
\usepackage[linesnumbered, ruled, vlined]{algorithm2e}

\title[Lightweight Exoplanet Simulation]{Extreme Learning Machines for Exoplanet Simulations: A Faster, Lightweight Alternative to Deep Learning}

\author[Tara P. A. Tahseen et al]{Tara P. A. Tahseen$^{1}$\thanks{E-mail: tara.tahseen.22@ucl.ac.uk (TPAT)},
Luís F. Simões$^{2}$,
Kai Hou Yip$^{1}$,
Nikolaos Nikolaou$^{1}$,
João M. Mendonça$^{4,5}$
\newauthor
and Ingo P. Waldmann$^{1}$
\\
$^{1}$Department of Physics and Astronomy, University College London, Gower Street, WC1E 6BT London, United Kingdom\\
$^{2}$ML Analytics, Lisbon, Portugal\\
$^{3}$Department of Physics and Astronomy, University of Southampton, Highfield, Southampton SO17 1BJ, UK \\
$^{4}$School of Ocean and Earth Science, University of Southampton, Southampton, SO14 3ZH, UK}

\date{Accepted XXX. Received YYY; in original form ZZZ}

\pubyear{\the\year{}}

\begin{document}
\label{firstpage}
\pagerange{\pageref{firstpage}--\pageref{lastpage}}
\maketitle

\begin{abstract}

Increasing resolution and coverage of astrophysical and climate data necessitates increasingly sophisticated models, often pushing the limits of computational feasibility. While emulation methods can reduce calculation costs, the neural architectures typically used—optimised via gradient descent—are themselves computationally expensive to train, particularly in terms of data generation requirements. This paper investigates the utility of Extreme Learning Machines (ELMs) as lightweight, non-gradient-based machine learning algorithms for accelerating complex physical models.

We evaluate ELM surrogate models in two test cases with different data structures: (i) sequentially-structured data, and (ii) image-structured data. For test case (i), where the number of samples $N \gg$ the dimensionality of input data $d$, ELMs achieve remarkable efficiency, offering a 100,000× faster training time and a 40× faster prediction speed compared to a Bi-Directional Recurrent Neural Network (BIRNN), whilst improving upon BIRNN test performance. For test case (ii), characterised by $d \gg N$ and image-based inputs, a single ELM was insufficient, but an ensemble of 50 individual ELM predictors achieves comparable accuracy to a benchmark Convolutional Neural Network (CNN), with a 16.4× reduction in training time, though costing a 6.9× increase in prediction time. We find different sample efficiency characteristics between the test cases: in test case (i) individual ELMs demonstrate superior sample efficiency, requiring only 0.28\% of the training dataset compared to the benchmark BIRNN, while in test case (ii) the ensemble approach requires 78\% of the data used by the CNN to achieve comparable results—representing a trade-off between sample efficiency and model complexity.

\end{abstract}


\section{Introduction}\label{sec:introduction}

Scientific inference in planetary, exoplanetary and Earth climate science heavily relies on simulations. Common uses of simulations include:
\begin{enumerate}
    \item \textbf{Forward models} used in Bayesian retrieval pipelines to constrain atmospheric parameters.
    \item \textbf{Global circulation models (GCMs)} employed to simulate planetary climate systems, including the Earth climate system.
    \item \textbf{Simulations of planetary host stars}, essential for understanding stellar signals.
\end{enumerate}
However, the computational expense associated with such simulations imposes significant limitations on the scope and scale of the science that can be performed. To address this, there is a pressing need for computationally cheaper alternatives that can retain sufficient accuracy for scientific analysis.

\subsection{Current Approaches to Reducing Computational Expense}
Efforts to accelerate simulations in exoplanet science often rely on scientifically motivated assumptions and model simplifications. For instance, in atmospheric modeling, the temperature-pressure (\(T-p\)) profile is often parametrized using simplified forms such as the \(N\)-point profile or the Guillot profile \citep{guillot_radiative_2010}, instead of simulating full radiative transfer. While such approaches can be computationally efficient, they are prone to biases—both known and unknown—which may compromise the robustness of the scientific conclusions drawn when these models are used in inference pipelines \citep{pluriel_strong_2020}.

\subsection{Surrogate Modeling as a Solution}
Surrogate modeling, or emulation, offers an alternative approach for accelerating simulations. This technique, though relatively nascent in exoplanet science, has been widely explored in adjacent fields such as Earth climate science, astrophysics, and broader physics and engineering domains. Surrogate models approximate the behavior of a physical or simulated system by using supervised machine learning algorithms trained on examples of input-output pairs from the target system.



Deep learning architectures, such as convolutional neural networks (CNNs), recurrent neural networks (RNNs), and graph neural networks (GNNs), have shown promise in creating high-precision surrogate models. However, the computational expense of training these models can itself become a bottleneck, particularly when complex architectures are involved. There is also a potential bottleneck in the time and computational resources required to synthesise a dataset to train and test such surrogate models, as this may likely involve many executions of the physical model which we would like to emulate. These factors limit the practical applicability of surrogate models which involve high training cost and high data-production cost, in exoplanet-related simulations.

A wealth of publications in astrophysics on surrogate modelling methodology has emerged in 2023 and 2024: these include applications in cosmology \citep{jense_complete_2024,  conceicao_fast_2024, rink_gravitational_2024}; galactic physics \citep{hirashima_surrogate_2023, hicks_galaxies_2024}; and exoplanet science \citep{himes_accurate_2022,  himes_towards_2023, unlu_reproducing_2023, tahseen_enhancing_2024}; to name a few.


A key commonality amongst the bulk of literature on surrogate modelling in astrophysics is the prevalence of neural architectures optimised using gradient descent. Gradient descent is a computationally intensive training procedure, often requiring graphics processing units (GPUs) and tensor processing units (TPUs). This computational demand limits applicability of such models to calculations which will be executed multiple times, and to researchers who have access to high-performance computing facilities. Engineering of such surrogate models is often time-consuming as the iteration process is bottle-necked by (i) the need to generate large numbers of training data via expensive simulations and (ii) the slow model training time. Model development time also poses an issue when retraining models to address model or data drift\footnote{The term \emph{data drift} refers to the systematic shift in the underlying distribution of the data from training to deployment. These changes in the distribution of data can cause the performance of trained models to deteriorate on new data and/or over time.}.

\subsection{Aim of This Study}
This study investigates whether the \textit{extreme learning machine} (ELM), which is fit through convex optimisation techniques like least squares regression rather than gradient descent, can serve as a viable alternative to the commonly-used deep learning algorithms for building surrogate models in exoplanet science. Practically, an ELM is a lightweight, computationally efficient machine learning algorithm. We benchmark the precision, computational efficiency and sample efficiency\footnote{By \emph{sample efficency} or {data efficiency} we refer to the capacity of a learning algorithm to produce good models at reduced cost in terms of number of training data.} of ELM surrogate models across two representative exoplanet-related simulations:
\begin{enumerate}
    \item \textbf{Atmospheric radiative transfer:} A low-dimensional, sequentially structured computation involving a high volume of training examples.
    \item \textbf{Stellar surface map to lightcurve conversion:} A high-dimensional image processing computation with a relatively low volume of training examples.
\end{enumerate}

These two cases were selected to enable a comprehensive evaluation of the ELM algorithm under differing conditions:
\begin{itemize}
    \item Experiment (i) features input data where the dimensionality (\(d\)) is much smaller than the number of training examples (\(n_{\text{train}}\)): \(d \ll n_{\text{train}}\); and input data has a sequential geometric structure.
    \item Experiment (ii) features input data where the dimensionality exceeds the number of training examples: \(d \gg n_{\text{train}}\); and input data has a 2D image structure.
\end{itemize}


\subsection{Summary of Contributions}
This work evaluates whether ELM can offer a practical alternative to more complex surrogate modeling approaches by balancing computational efficiency with predictive accuracy. By addressing these questions, we aim to expand the toolbox available for computationally affordable exoplanet science.

\section{Experiments}\label{sec:experiments}

\subsection{Benchmark Deep Learning Models}

The deep learning models which we use as benchmarks for which to compare the ELM performance are chosen for being the state-of-the-art for the two different experiments we describe in Section \ref{sec:expt-1} and Section \ref{sec:expt-2}. More details on the selection process of these benchmark models can be found in Section \ref{sec:selection-of-benchmark-models}.

In addition to these state-of-the-art benchmark models, we also use a three-layer Dense Neural Network (DNN) as an additional point of comparison in each experiment. Architectural details of the DNNs can be found in Section \ref{sec:dnn}.

\subsection{\textit{Experiment 1}: Emulation of Numerical Radiative Transfer Schema}\label{sec:expt-1}

This experiment involves an investigation into the suitability of the ELM as a candidate algorithm with which to produce surrogate models for integration within climate models. For this investigation, we select a task of emulating short-wave two-stream radiative transfer; a notoriously complex and slow component in atmospheric simulations. We perform grid optimisation over ELM parameters (this is described in more detail in Section \ref{sec:results-expt-1}), and evaluated the grid of ELMs in terms of (1) accuracy against a validation set; (2) training time; (3) prediction time. The candidate ELM model with highest accuracy on the validation set is evaluated against a test set, and compared to a benchmark model, a Bi-directional Recurrent Neural Network (BIRNN). For the combination of hyperparameters which produces the best performing model, we then conduct an investigation into how choice of training set size and random seed with which to initialise network weights affects test loss.

The data and codes used in this experiment are the same as those detailed in Section 2 of \cite{tahseen_enhancing_2024}. In brief, data used for this work comprises 1,000 epochs of simulation runtime using the \texttt{OASIS} global circulation model to simulate the Venus atmosphere \citep{mendonca_modelling_2020}, using a two-stream radiative transfer model as detailed in \cite{mendonca_new_2015}. Each epoch consists of 10,242 samples -- vertical profiles of atmospheric variables across 49 layers ($n_\text{layers}$) with 50 boundary levels ($n_\text{levels}$).

\subsubsection{Data Preprocessing}\label{sec:expt-1-data-preprocessing}

\begin{figure}[!ht]
    \centering    \includegraphics[width=1\linewidth]{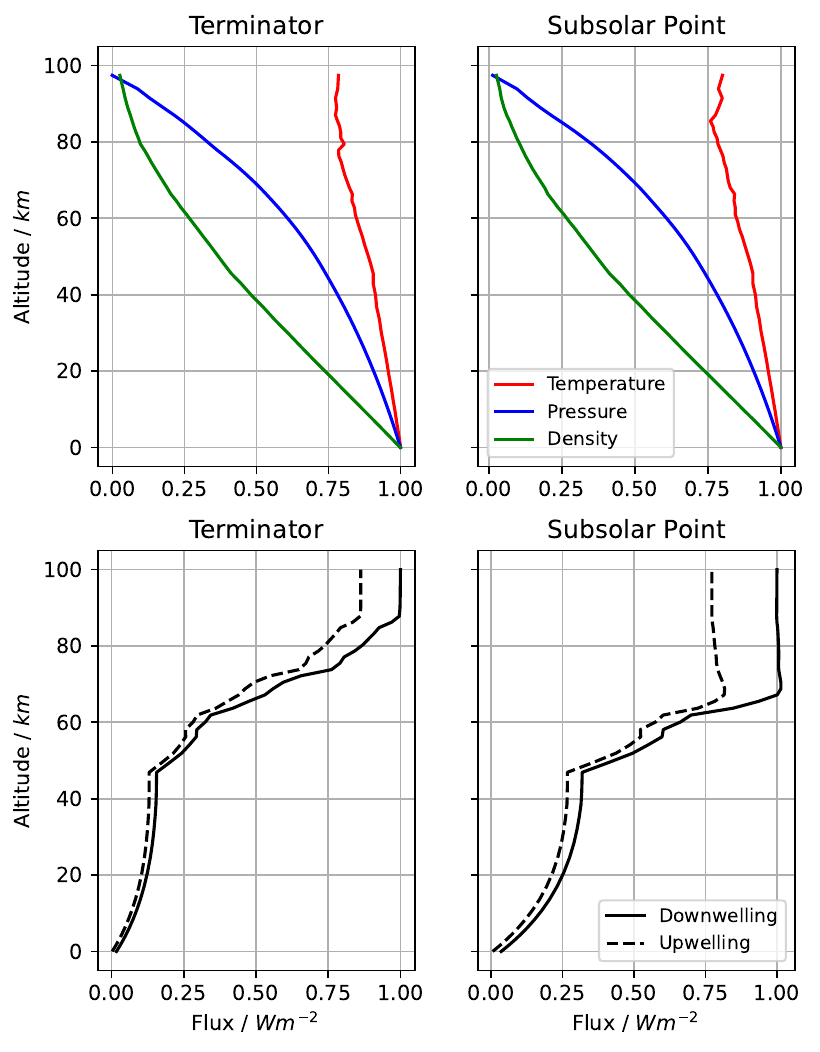}
    \caption{Examples of input (top) and target (bottom) data used for training surrogate models in Experiment 1. The data corresponds to (left) an atmospheric column at the day-night terminator and (right) an atmospheric column at the subsolar point. The input and target data have been preprocessed from their raw values, as detailed in Section \ref{sec:expt-1-data-preprocessing}. Specifically, temperature and pressure are log-transformed and normalized by their surface log values, while gas density is rescaled by raising it to the power of 0.25 and then normalized by the surface gas density to the power of 0.25.}
    \label{fig:experiment-1-input-target}
\end{figure}

Data were preprocessed according to Section 3.2 of \cite{tahseen_enhancing_2024}. In this work, the test set comprised 768,000 samples, the validation set comprised 768,000 samples, and the train set comprised 3,584,000 samples. Input data were reshaped to have dimensions $(n_{\text{samples}}, \, n_{\text{input features}} * n_{\text{layers}})$, and output data were reshaped to have dimensions $(n_{\text{samples}}, \, n_{\text{output features}} * n_{\text{levels}})$. Input features consist of atmospheric profiles of temperature, pressure, and gas density across atmospheric layers. Output variables consist of downwelling short-wave flux and upwelling short-wave flux across atmospheric levels. Examples of preprocessed input and target data are displayed in Figure \ref{fig:experiment-1-input-target}.

\subsubsection{Benchmark Surrogate Model}

The benchmark surrogate model for the two-stream radiative transfer physical model was a BIRNN, taking vector input profiles of scaled temperature $\textbf{T}$, scaled pressure $\textbf{p}$, and scaled gas density $\bm{\rho}$ along a given atmospheric column, along with scalar variables including the cosine of the solar zenith angle $\mu$, and surface albedo $\alpha$, associated with the specific column. The outputs of the BIRNN are the scaled upwelling and downwelling flux profiles, $\bm{F^{\uparrow}}$ and $\bm{F^{\downarrow}}$, for layer boundaries along the column. The methods of scaling the vector input and target variables are detailed in Section 3.2 of \cite{tahseen_enhancing_2024}. The benchmark model was trained using the full train set comprising 3,584,000 samples. The batch size used for training was 512 samples, and the BIRNN was trained with an MSE loss.

\subsection{\textit{Experiment 2}: Emulation of Stellar Surface Map to Lightcurve Model}\label{sec:expt-2}

This experiment involved producing an ELM surrogate model for generating a light curve from an input stellar surface map, and followed the same training and evaluation strategy as outlined for Experiment 1 in Section \ref{sec:expt-1}. The dataset contains simulated data products of V1298 using \texttt{PAStar} \citep{petralia_pastar_2024}. The input data consists of a map showing longitude and altitude of the star’s photospheric surface. It is a boolean map where 1 indicates the presence of stellar spots and 0 indicates the photosphere of the star. The output is a light curve time series documenting the changes in flux as the star rotates from the observer’s point of view. In this toy example, we did not simulate faculae and set the temperature contrast to 1000K.


\begin{figure}[!ht]
    \centering    \includegraphics[width=1\linewidth]{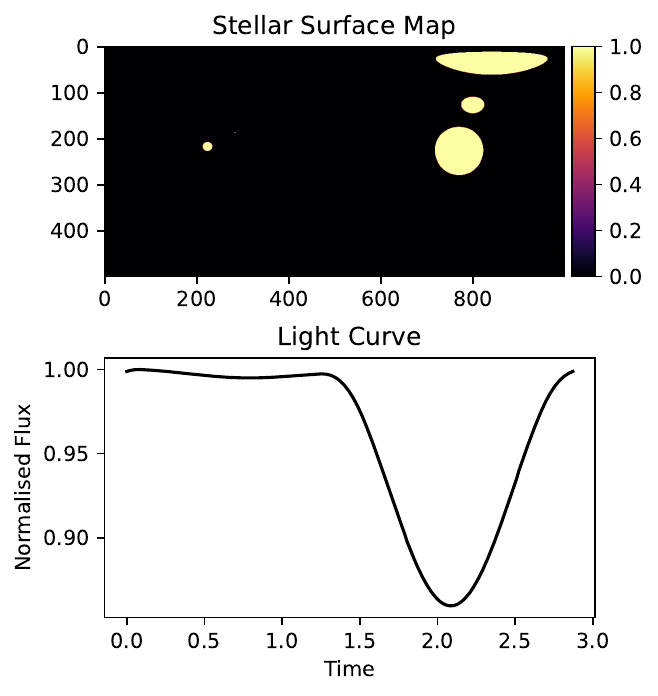}
    \caption{Example of an input (top) target (bottom) pair for training surrogate models in Experiment 2.}
    \label{fig:experiment-2-input-target}
\end{figure}

\subsubsection{Data Preprocessing}
Data were randomly shuffled and split into the train set, validation set and test set, which comprised 6400, 1600, and 2000 samples respectively. Input data was of dimension (999, 499) per sample, and target data was of dimension (200,). For use with the ELM models, input data was flattened to dimension (498501,). Target data was standardised by removing the mean and scaling to unit variance, whereby the mean and the variance were calculated from the train set only.

\subsubsection{Benchmark Surrogate Model}

The benchmark surrogate model for Experiment 2 was a CNN with architecture and trainable parameters detailed in Table \ref{tab:cnn-architecture}. The inputs to the model are stellar surface maps of shape $(999, 499)$, and outputs are light curves of shape $(200,)$, with an example of an input-target pair displayed in Figure \ref{fig:experiment-2-input-target}. The batch size used for training was 64, and the CNN was trained with an MAE loss.

\begin{table}[h!]
    \centering
    \begin{tabular}{ccc}\hline
         \textbf{CNN Layer}&  \textbf{Output Shape}& \textbf{Number of Parameters}\\\hline
         Input&  $(n_\text{samples}, 999, 499)$& 0\\
         Reshape&  $(n_\text{samples}, 999, 499, 1)$& 0\\
         2D Convolution&  $(n_\text{samples}, 999, 499, 4)$& 40\\
         2D Convolution&  $(n_\text{samples}, 999, 499, 4)$& 148\\
         2D Max Pooling&  $(n_\text{samples}, 999, 249, 4)$& 0\\
         2D Convolution&  $(n_\text{samples}, 999, 249, 8)$& 296\\
         2D Convolution&  $(n_\text{samples}, 999, 249, 8)$& 584\\
         Flatten&  $(n_\text{samples}, 247008)$& 0\\
         Dense&  $(n_\text{samples}, 200)$& 49,401,800\\\hline
 \multicolumn{2}{c}{\textbf{Total Parameters}}&49,402,868\\\hline
    \end{tabular}
    \caption{Architecture and number of trainable parameters of the CNN benchmark surrogate model for Experiment 2.}
    \label{tab:cnn-architecture}
\end{table}

\section{Methods}\label{sec:methods}

The extreme learning machine technique employed in this paper differs from commonly-used deep learning paradigms in two fundamental ways: (1) the hidden layer weights are randomly initialised and remain fixed throughout training; (2) only the output layer weights are optimised, which is accomplished through a direct analytical solution rather than iterative gradient descent. This optimisation approach greatly reduces the computational expense of model training compared to deep learning methods. We describe the mathematics of the ELM below.

\subsection{Extreme Learning Machines}





ELMs \citep[mathematically formalised and described in][]{schmidt_feedforward_1992, pao_learning_1994, huang_extreme_2004, nan-ying_liang_fast_2006, huang_extreme_2006, deng_regularized_2009} can be considered a subset of neural networks, but differ from traditionally back-propagation-based trained neural networks in their parameter initialisation, whereby parameters controlling the internal network state are untrainable, enabling a gradient-free method of parameter fitting for parameters mapping from the internal network state to the output (as described mathematically in sections below). \cite{mcdonnell_fast_2015, huang_trends_2015} discuss applications of ELMs across different domains. 

\subsubsection{Architecture}\label{architecture}

An ELM maps from an input vector $\mathbf {X} \in \mathbb{R}^{N \times d}$ to an output vector $\mathbf {Y} \in \mathbb{R}^{N \times m}$, where $N$ is the number of individual samples inputted to the model, $d$ is the dimensionality of a given sample, and $m$ is the dimensionality of the model output. For the experiments in this work, we consider the values of $N_\text{train}, d, m$ as outlined in Table \ref{tab:nmd}, where $N_\text{train}$ is the number of samples in the training dataset. The mathematics of the ELM are as follows:

\begin{table}
    \centering
    \begin{tabular}{|c|c|c|c|}
       Experiment  & $N_\text{train}$ & $d$ & $m$\\\hline
        1 & 10,000 & 147 & 100\\
        2 & 6,400 & 498,501 & 200 \\
    \end{tabular}
    \caption{The number of samples $N_\text{train}$, sample input dimensionality $d$, and sample output dimensionality $m$, which are considered across Experiment 1 and Experiment 2. Experiments were chosen to cover cases of both $N_\text{train}>>d$, as in Experiment 1, and $N_\text{train}<<d$, as in Experiment 2.}
    \label{tab:nmd}
\end{table}


We define a fully-connected hidden layer $h$ with $L$ hidden neurons, and a weights matrix 
\begin{equation}
    \mathbf{W} \in \mathbb{R}^{L \times d}
    \label{eq:W}
\end{equation}
The hidden layer output matrix $\mathbf {H} \in \mathbb{R}^{N \times L}$ can then be written as follows: 
\begin{equation}
    \mathbf{H} = g(\mathbf {X} \mathbf{W}^T + \mathbf{b})
    \label{eq:H}
\end{equation}
where $\mathbf{b} \in \mathbb{R}^{L}$ is a biases matrix and $g$ is a non-linear activation function.

The output vector $Y$ is then formulated as follows: 
\begin{equation}
    \mathbf {Y} = \mathbf {H} \bm{\beta}
    \label{eq:Y}
\end{equation}
where $\bm{\beta} \in \mathbb{R}^{L \times m}$.

\subsubsection{Parameter Fitting}\label{sec:elm-parameter-fitting}
The ELM described in Section \ref{architecture} is parametrised by $\mathbf{W}$, $\mathbf{b}$ and $\bm{\beta}$. $\mathbf{W}$ and $\mathbf{b}$ are randomly initialised, and $\bm{\beta}$ is generated using training examples. Given a training set consisting of inputs $\hat{\mathbf {X}} \in \mathbb{R}^{N_{\text{train}} \times d}$ and targets $\hat{\mathbf {Y}} \in \mathbb{R}^{N_{\text{train}} \times m}$, we construct $\bm{\beta}$ as
\begin{equation}
    \bm{\beta} = \mathbf {H}(\hat{\mathbf {X}})^\dagger \hat{\mathbf {Y}}
    \label{eq:training}
\end{equation}
where $\mathbf {H}^\dagger$ is the Moore-Penrose pseudo-inverse of $\mathbf {H}$. 


In this work, we compute $\bm{\beta}$ as follows using a Tikhonov regularisation approach:
\begin{equation}
{\boldsymbol {\beta }}= ( {\mathbf {H}}^{T} {\mathbf {H}} + 
\alpha \mathbf{I})^{-1}( {\mathbf {H}}^{T} {\hat{\mathbf {Y}}})
\end{equation}
where $\alpha=10^{-9}$ is a fixed regularisation parameter.

\subsubsection{Hyperparameter Sensitivity Investigation}




This study explored the sensitivity of ELMs to various hyperparameters, focusing on seven activation functions—sigmoid, softsign, tanh, LeCun tanh, softplus, rectifier, and Gaussian. Detailed descriptions of these functions are provided in Section \ref{appendix:activation-functions}. Additionally, the hidden layer weights were randomly initialized using either a normal or a uniform distribution, as described in Section \ref{appendix:distributions}. For a deeper understanding of the practical implementation of ELMs in high-performance big data applications, readers are referred to \cite{akusok_high-performance_2015}.

\subsection{Ensemble Predictors}


This work also utilises ensemble predictors. When using such ensembles, all individual predictors were trained using the same input data and target data, and in the case of generating predictions, model predictions were generated across a \texttt{for} loop of all predictors and aggregated together by taking the mean of all predictions. This process is described in Algorithm \ref{alg:ensemble_prediction}. Diversity was introduced among the ensemble members by varying the random seed used to initialise hidden layer weights.

       




\begin{algorithm}
\caption{Prediction Execution for an Ensemble Model}
\label{alg:ensemble_prediction}
\KwIn{
    \begin{itemize}
        \item A dataset $X$ with $n$ samples and $d$ features
        \item $L$ individual predictors $\{P_1, P_2, \dots, P_L\}$
    \end{itemize}
}
\KwOut{Aggregated predictions $\hat{y}$ for all $n$ samples}

\For{$i = 1$ \KwTo $L$}{
    Execute predictions for predictor $P_i$ on dataset $X$;
    Store predictions as vector $\hat{y}_i$;
}
$\hat{y} \gets \frac{1}{L} \sum_{i=1}^L \hat{y}_i$;

\Return{$\hat{y}$}
\end{algorithm}

\subsection{Dense Neural Networks}\label{sec:dnn}

In both experiments, a Dense Neural Network (DNN) is included as a baseline to provide a standard and broadly interpretable point of comparison alongside the ELM, BIRNN, and CNN models. While DNNs are not the primary focus of this study, they provide a useful reference point given their generality and computational efficiency. Their inclusion enables a more comprehensive assessment of the ELM’s performance relative to three widely used neural architectures: BIRNNs, CNNs, and DNNs. For consistency, the same DNN architecture was used in both experiments, despite the fundamentally different nature of the data -- low-dimensional sequential inputs in Experiment 1 and high-dimensional image inputs in Experiment 2.

To manage overfitting risks in the context of limited data in Experiment 2, the DNN architecture was kept deliberately simple, balancing parameter count and model capacity. With regard to Experiment 1, the design choice builds on findings from \citep{ukkonen_exploring_2022}, where a three-layer DNN with 128 neurons per layer showed inferior performance compared to a BIRNN, but nonetheless established a strong baseline. In this work, we adopt a similar 3-layer DNN structure, but increase the number of neurons in the first two layers to 512 and 256 respectively (followed by 128), with the aim of improving model expressivity and potentially narrowing the performance gap with the more specialised BIRNN. Here, we make no claim that our DNNs are fully optimised, but rather include them as a standard and broadly interpretable point of comparison.

\subsubsection{Experiment 1}
The input to the network was a 147-dimensional feature vector. The model architecture consisted of three hidden dense layers followed by an output layer:

\begin{itemize}
    \item Dense layer with 512 units (75,776 parameters)
    \item Dense layer with 256 units (131,328 parameters)
    \item Dense layer with 128 units (32,896 parameters)
    \item Output layer with 100 units (12,900 parameters)
    \item The total number of trainable parameters was 252,900.
\end{itemize}

All hidden layers employed the ReLU activation function, and the output layer used a linear activation function. L2 regularization $(\lambda = 10^{-4})$ was applied to all dense layers. The model was trained for up to 50 epochs using the Adam optimizer with a batch size of 1024 and early stopping with a patience of 10 epochs. The loss function was MSE, in alignment with the evaluation metrics of benchmark models for this task. The DNN was trained using the same training set used for the BIRNN, with 3,584,000 samples.


\subsubsection{Experiment 2}

This experiment used a higher-dimensional input space, a 498,501-dimensional feature vector, and the model consisted of:

\begin{itemize}
    \item Dense layer with 512 units (255,233,024 parameters)
    \item Dense layer with 256 units (131,328 parameters)
    \item Dense layer with 128 units (32,896 parameters)
    \item Output layer with 200 units (25,800 parameters)
    \item The total number of trainable parameters was 255,423,048.
\end{itemize}

As with Experiment 1, a ReLU activation function was used for all hidden layers, with a linear activation at the output. L2 regularization $( \lambda = 10^{-4})$ was applied to all dense layers. The model was trained using the Adam optimizer for up to 50 epochs, with early stopping (patience = 10) and a batch size of 8 due to the larger input data size. The loss function was MAE, selected to match the criteria used in benchmark deep learning models for this task. The DNN was trained with the same training set as for the CNN, with 6,400 samples.

\subsection{Performance Metrics}\label{sec:performance-metrics}

To evaluate the performance of models used in this work, we employed the following metrics:
\begin{enumerate}
    \item \textbf{Model Training Time ($T_\text{train}$)}: this time is measured in seconds on a single CPU core.\\
    \item \textbf{Model Prediction Time ($T_\text{predict}$)}: this time is measured in seconds on a single CPU core.\\
    \item \textbf{Mean Squared Error (MSE)}:
    \begin{equation}\label{equation:mse}
        \text{MSE} = \frac{1}{n} \sum_{i=1}^n \left( y_i - \hat{y}_i \right)^2
    \end{equation}

    \item \textbf{Mean Absolute Error (MAE)}:
    \begin{equation}\label{equation:mae}
        \text{MAE} = \frac{1}{n} \sum_{i=1}^n \left| y_i - \hat{y}_i \right|
    \end{equation}

    \item \textbf{Mean Absolute Percentage Error (MAPE)}: 
    \begin{equation}\label{equation:mape}
        \text{MAPE} = \frac{1}{n} \sum_{i=1}^n \left| \frac{y_i - \hat{y}_i}{y_i} \right| \times 100
    \end{equation}

\end{enumerate}
where $\hat{y}_i$ is the predicted output and $y_i$ is the target, for the $i$-th sample in a dataset of size $n$ in Equations \ref{equation:mse}, \ref{equation:mae} and \ref{equation:mape}.


\section{Results}\label{sec:results}

\subsection{\textit{Experiment 1}}\label{sec:results-expt-1}

Multiple ELMs were trained across a grid search of activation function ($\alpha$), corresponding to all of those described in Section \ref{appendix:activation-functions}, and hidden layer sizes ($n_\text{hidden}$) between 150 and 1,000 neurons, with weights randomly initialised (using a fixed random seed) using two different distributions ($f$), normal and uniform, as formulated in Section \ref{appendix:distributions}. ELMs across the grid search were evaluated primarily using MSE, with MAE and MAPE also being calculated, as shown in Figure \ref{fig:expt-1-grid-search}. The plots in Figure \ref{fig:expt-1-grid-search} demonstrate consistency across these three metrics. The 'best-performing' ELM was chosen as the ELM which minimised the MSE on the validation set, and corresponded to a hidden layer with 1,000 neurons, a Gaussian activation function, and weights randomly initialised using a normal distribution. ELMs in the grid search were trained using a randomly sampled subset of 10,000 samples from the training dataset -- this was chosen in the interest of minimising ELM training time.

\begin{figure}[!ht]
    \centering    \includegraphics[width=1\linewidth]{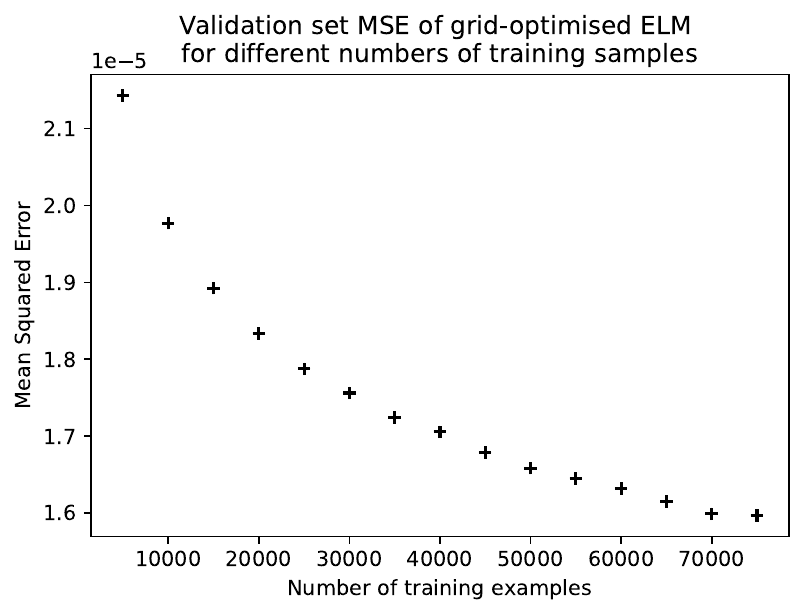}
    \caption{Mean squared error on the validation set for ELM models with a fixed set of hyperparameters corresponding to those of the best performing model within the grid search, but for variable number of training samples, $n_\text{train}$.}
    \label{fig:expt-1-sample-efficiency}
\end{figure}

\begin{figure}[!ht]
    \centering    \includegraphics[width=1\linewidth]{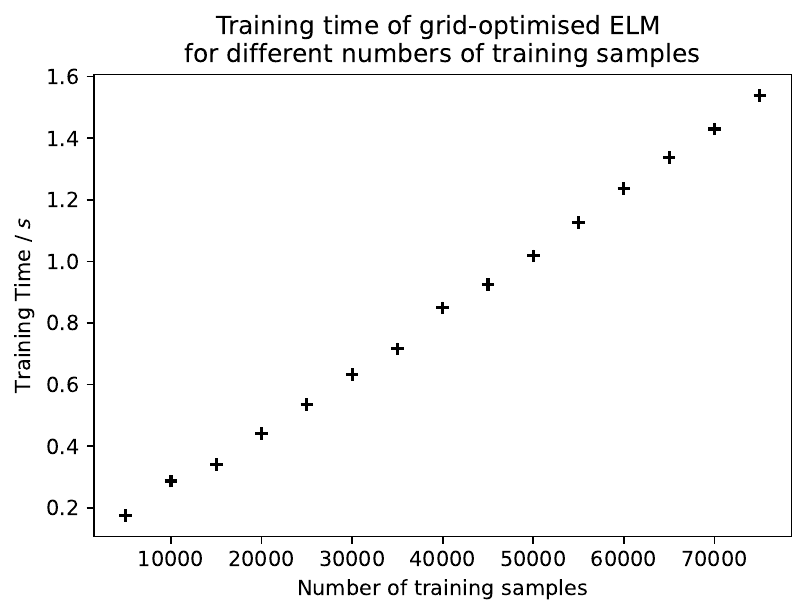}
    \caption{Training time of ELM models with a fixed set of hyperparameters corresponding to those of the best performing model within the grid search, but for variable number of training samples, $n_\text{train}$. We observe a roughly linear dependence of the training time on $n_\text{train}$.}
    \label{fig:expt-1-training-time-vs-samples}
\end{figure}

For the best-performing ELM, experiments were then conducted varying the number of training samples and recording the effect on the MSE of the validation set, as shown in Figure \ref{fig:expt-1-sample-efficiency}, and on the training time, as shown in Figure \ref{fig:expt-1-training-time-vs-samples}. Performance stability as a function of the random seed used for weight initialisation was investigated for the best-performing ELM, for which the resulting plot is displayed in Figure \ref{fig:expt-1-grid-search-2}.

\newpage
\subsection{\textit{Experiment 2}}


\subsubsection{ELMs}\label{sec:experiment-2-elms}

As described for Experiment 1 in Section \ref{sec:results-expt-1}, multiple ELMs were trained across a grid of hyperparameters with variable hidden size ($n_\text{hidden}$), activation function ($\alpha$), and distribution ($f$) from which to randomly sample weights. The combination of hyperparameters which optimised the MSE of validation set predictions was $n_\text{hidden}=2,500$, $\alpha=\text{softplus}$, and $f=\text{uniform}$. The MSE, MAE and MAPE of models across this grid search are displayed in Figure \ref{fig:expt-2-grid-search}; we find that the grid-optimised ELM achieved test performances around an order of magnitude worse than that of the CNN, across MSE, MAE, and MAPE, as shown in Table \ref{tab:results-summary-experiment-2}.




\subsubsection{Ensemble Predictors}

As no ELM within the grid search performed comparably with or better than the CNN, we proceeded to experiment with using Ensemble ELMs. Ensembles were trained consisting of individual predictors with fixed hyperparameters corresponding to those optimised in Section \ref{sec:experiment-2-elms}, but with variable random seed governing weights initialisation across individual predictors.

An ensemble of 50 predictors was trained which outperformed all individual ELMs, and performed with a 1.2-1.3 times reduction in MSE, MAE and MAPE compared to the CNN. 


To complement the main analysis, we conducted a separate study comparing CNN and ELM ensembles under matched conditions. Rather than comparing the CNN ensemble to the full 50-predictor ELM ensemble, we limited both ensemble types to a maximum of 15 predictors to ensure a fair and computationally feasible comparison. This analysis focuses on how test performance varies with ensemble size, providing insight into the relative ensembling behaviour of CNNs and ELMs. Like the ELM ensemble, we diversify the predictors within the CNN ensemble by varying the random seed used to initialise
weights (prior to optimisation).

\begin{figure}[!ht]
    \centering    \includegraphics[width=1\linewidth]{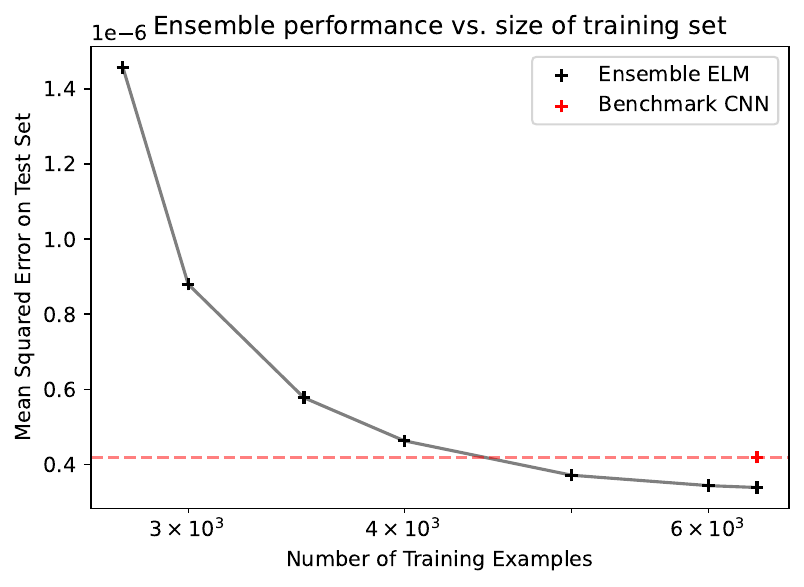}
    \caption{Mean squared error on the test set of ELM ensemble models trained using training sets of varying sizes. The MSE on the test set of the benchmark CNN trained using the full training set is also plotted for comparison (note that the dashed line is plotted irrespective of the values on the horizontal axis, and corresponds to 6,400 training samples).}
    \label{fig:experiment-2-sample-efficiency}
\end{figure}


\subsection{Results Summary}

\begin{figure*}[!ht]
    \centering    \includegraphics[width=1\linewidth]{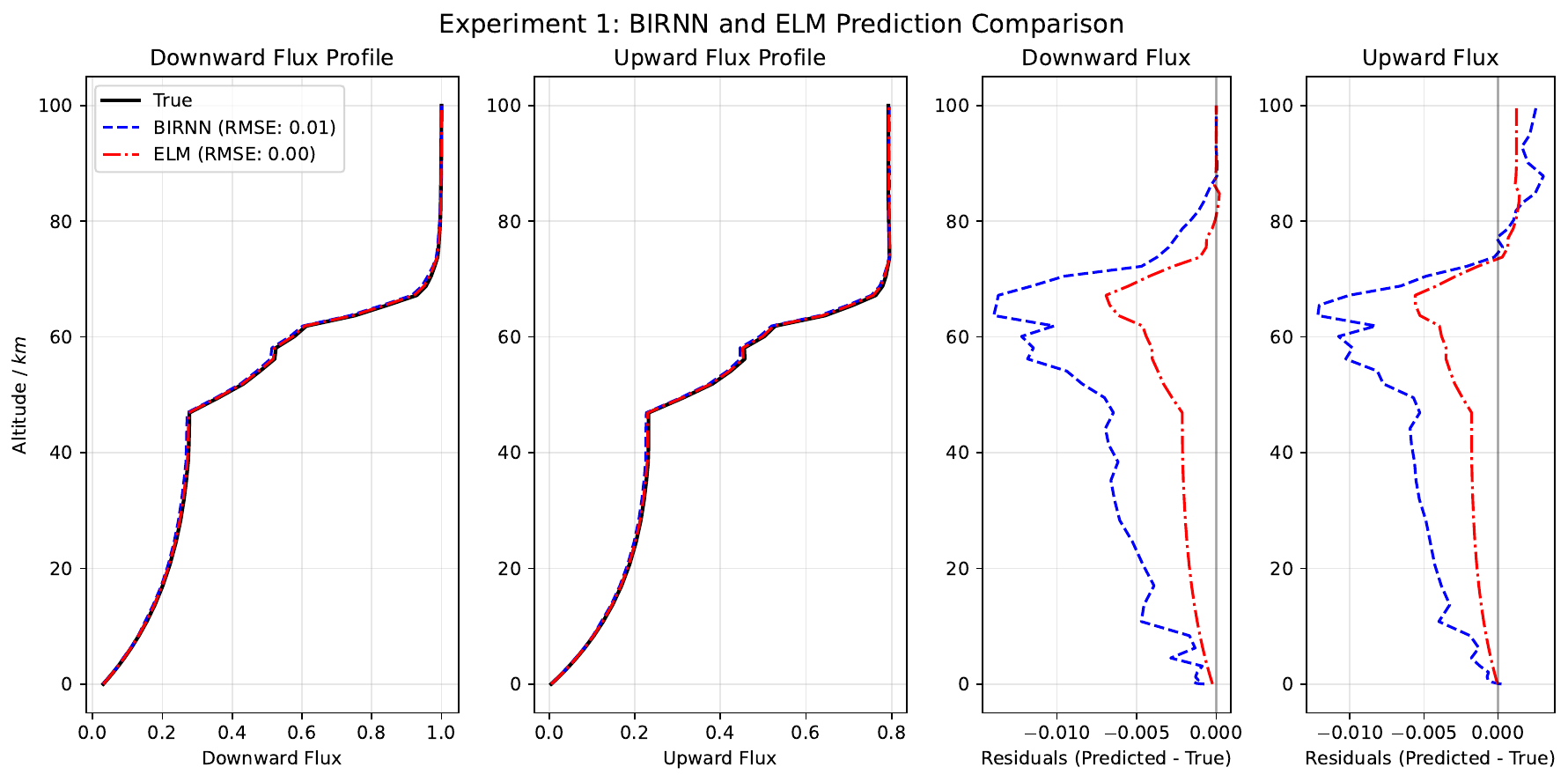}
    \caption{BIRNN and ELM predictions for downwelling flux (leftmost) and upwelling flux (second from left) for a test sample, using the grid-optimised ELM for Experiment 1. The test sample was chosen as being that with median ELM prediction error (calculated as the RMSE) out of the test set. Residuals are plotted for downwelling flux (second from right) and upwelling flux (rightmost). The residuals for this specific test sample are shown as being smaller in magnitude for the ELM prediction than the BIRNN prediction across all altitude levels for both target quantities.}
    \label{fig:experiment-1-example-prediction}
\end{figure*}

\begin{figure*}[!ht]
    \centering    \includegraphics[width=.8\linewidth]{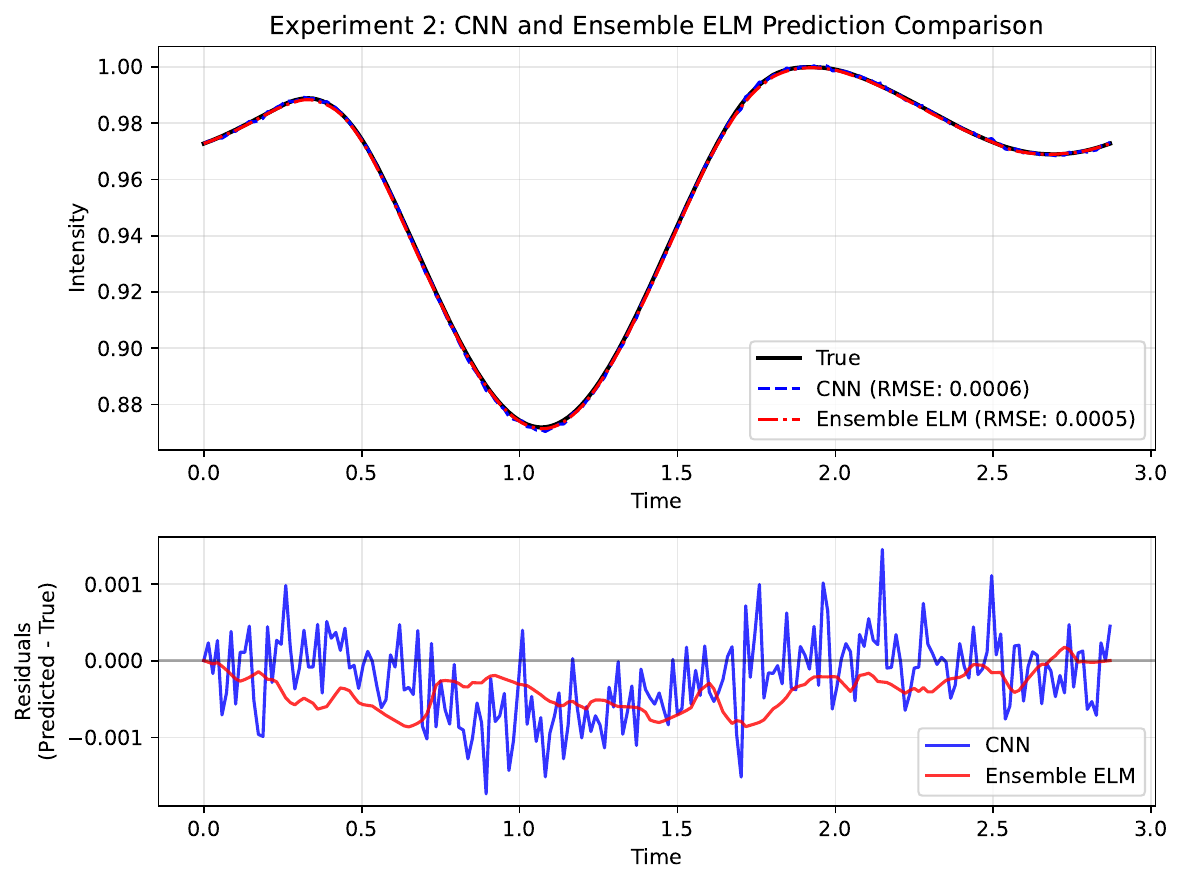}
    \caption{CNN and Ensemble ELM predictions for a test sample (top), using the optimal ELM-based model in Experiment 2. Residuals for the CNN and Ensemble ELM compared to the target lightcurve (bottom) display a smoother curve for the Ensemble ELM compared to the CNN, and also illustrate an underprediction of the Ensemble ELM across most of the lightcurve, which we do not see with the CNN.}
    \label{fig:experiment-2-example-prediction}
\end{figure*}

\begin{figure*}[!ht]
    \centering    \includegraphics[width=1\linewidth]{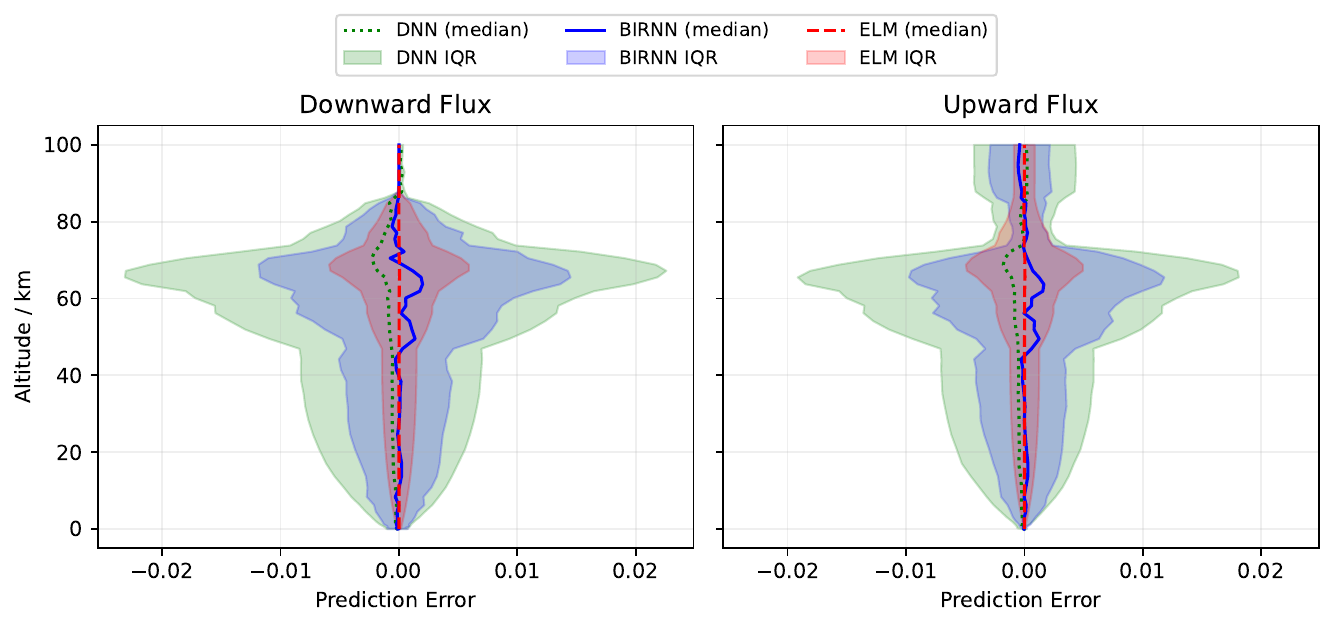}
    \caption{Median prediction error with Inter-Quartile Range (IQR) for DNN, BIRNN and ELM for Experiment 1. Test performance is displayed for the two targets: (left) downwelling flux, (right) upwelling flux, across the altitudinal extent of the simulation. The prediction error is computed as the MSE between the model prediction and target value.}
    \label{fig:expt-1-ffnn}
\end{figure*}

\begin{figure*}[!ht]
    \centering    \includegraphics[width=1\linewidth]{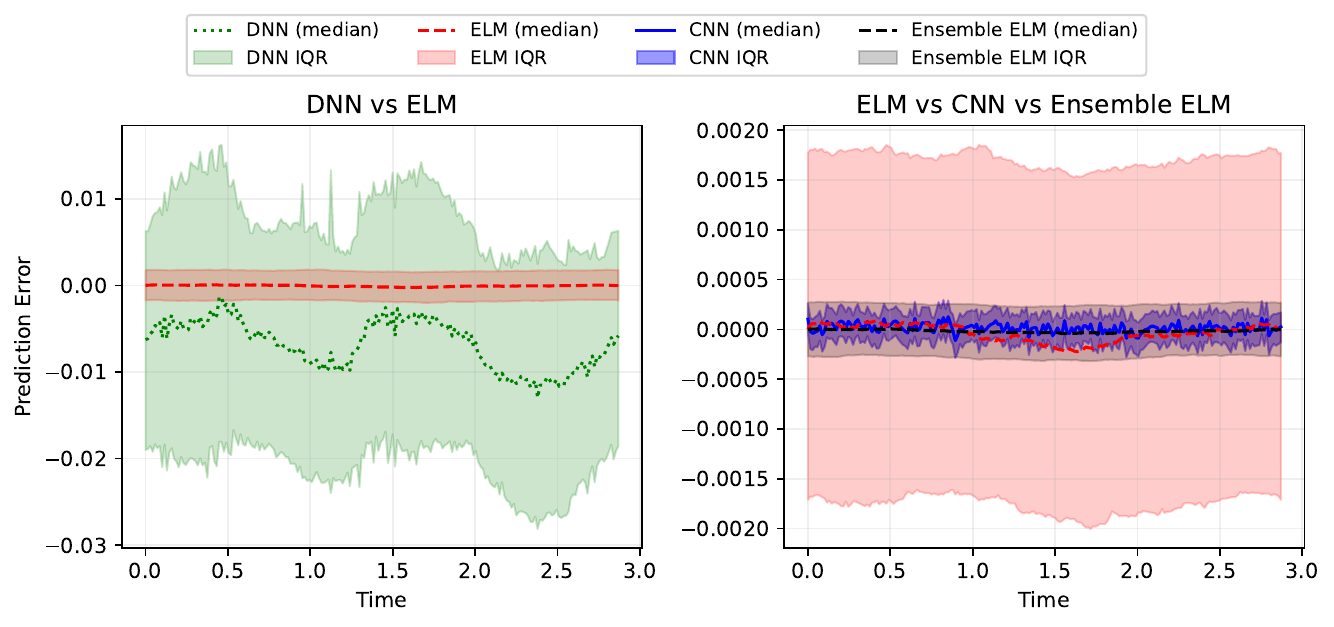}
    \caption{Median prediction error with Inter-Quartile Range (IQR) for DNN, CNN, ELM and Ensemble ELM for Experiment 2. Test performance is displayed across all timepoints of the target lightcurve. The prediction error is computed as the MSE between the model prediction and target value. We display prediction error distributions for the 4 models across two different plots for visual clarity: (left) DNN and ELM, (right) ELM, CNN, Ensemble ELM.}
    \label{fig:expt-2-ffnn}
\end{figure*}

\begin{figure}[!ht]
    \centering    \includegraphics[width=1\linewidth]{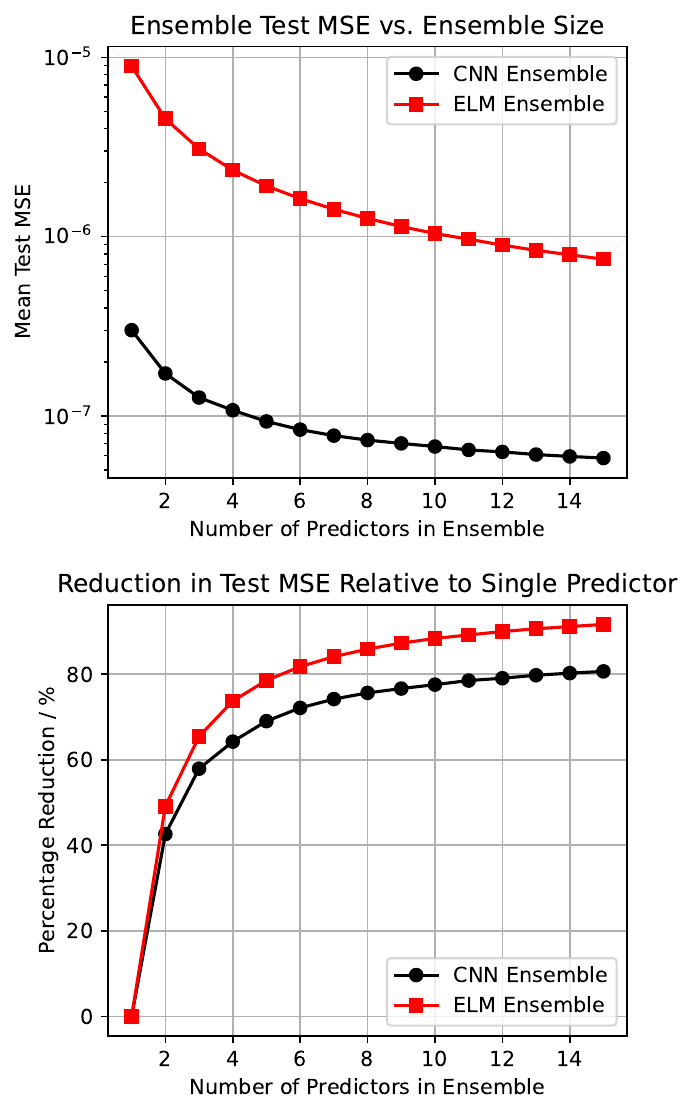}
    \caption{\textbf{Top}: Mean test MSE plotted on a logarithmic scale versus the number of predictors in the ensemble. The CNN ensemble (black circles) consistently achieves lower MSE than the ELM ensemble (red squares), with both showing improved performance as ensemble size increases.
    \textbf{Bottom}: Percentage reduction in test MSE relative to a single predictor for both CNN and ELM ensembles. Both methods benefit from ensembling and display a similar trend, with the ELM ensemble showing a steeper initial improvement compared to the CNN ensemble.}
    \label{fig:ensemble-performance}
\end{figure}

The summary of key results comparing the benchmark deep learning models with the best performing ELM or ELM ensemble is displayed in Table \ref{tab:results-summary-experiment-1} for Experiment 1 and Table \ref{tab:results-summary-experiment-2} for Experiment 2. An example of the median model performance for the best performing ELM-based model is displayed in Figure \ref{fig:experiment-1-example-prediction} for Experiment 1 and Figure \ref{fig:experiment-2-example-prediction} for Experiment 2; further plots characterising error distributions of all models across the test sets are displayed in Figure \ref{fig:expt-1-ffnn} for Experiment 1 and Figure \ref{fig:expt-2-ffnn} for Experiment 2. Comparison between test performance of ELM ensembles and CNN ensembles for Experiment 2 is displayed in Figure \ref{fig:ensemble-performance}.


\begin{table*}[h!]
    \centering
    \begin{tabular}{|c|c|c|c|c|} \hline 
         &  \textbf{Number of Neurons in Hidden Layer}&  \textbf{Activation Function}& \textbf{Distribution} &\textbf{Subset of Train Set Used for Fitting} \\ \hline 
         \textbf{Experiment 1}&  1,000&  Gaussian& Normal &10,000 (0.28\%)\\ \hline 
         \textbf{Experiment 2}&  2,500&  Softplus& Uniform &6,400 (100\%)\\ \hline
    \end{tabular}
    \caption{Hyperparameters of the best performing ELMs across hyperparameter grid searches for Experiment 1 and Experiment 2.}
    \label{tab:optimal-hyperparameters-table}
\end{table*}

\begin{table*}[h!]
    \centering
    \begin{tabular}{|c|c|c|c|} \hline 
         &  \textbf{Benchmark BIRNN} & \textbf{DNN}& \textbf{Best ELM}\\ \hline 
         \textbf{Model Training Time ($T_\text{train}$) / $\text{s}$}&  146,716.66 &3,502.3&\textbf{0.45}\\ \hline 
         \textbf{Model Prediction Time ($T_\text{predict}$) / $\text{s}$}&  41.78 &1.39& \textbf{1.14}\\ \hline 
         \textbf{Mean Squared Error (MSE)}&  $1.28 \times 10^{-4}$ &$2.27 \times 10^{-4}$& $\mathbf{(2.00 \pm 0.01) \times 10^{-5}}$\\ \hline 
         \textbf{Mean Absolute Error (MAE)}&  $6.00 \times 10^{-3}$ &$9.25 \times 10^{-3}$& $\mathbf{(2.37 \pm 0.01) \times 10^{-3}}$\\ \hline
 \textbf{Mean Absolute Percentage Error (MAPE) / \%}& 2.34 &3.41&$\mathbf{0.842 \pm 0.003}$\\\hline
    \end{tabular}
    \caption{Performance metrics of the benchmark BIRNN, DNN and the grid-optimised ELM trained as a surrogate model for Experiment 1. Prediction time $T_\text{predict}$ was measured using the validation set of 768,000 samples. Errors in the MSE, MAE and MAPE for the best ELM have been calculated as the standard deviation across different random initialisations of the ELM weights, corresponding to the top plot in Figure \ref{fig:expt-1-grid-search-2}. The best ELM used a hidden layer with 1,000 neurons, a Gaussian activation function, and sampled the hidden weights from a normal distribution; this ELM was trained using a subset of the full training set, using 10,000 samples, whilst the benchmark model was trained using the full train set of 3,584,000 samples. For each metric, the best value achieved across the benchmark BIRNN, DNN, and the best ELM is highlighted in bold. The batch sizes used for making predictions across models here was 4,096.}
    \label{tab:results-summary-experiment-1}
\end{table*}

\begin{table*}[h!]\label{table:results-expt-2}
    \centering
    \begin{tabular}{|c|c|c|c|c|} \hline 
         &  \textbf{Benchmark CNN} &\textbf{DNN}& \textbf{Best ELM} &\textbf{Best ELM Ensemble}\\ \hline 
         \textbf{Model Training Time ($T_\text{train}$) / $\text{s}$}&  151,413.6 &7,475.1&  $\mathbf{211.0}$&9,216.2\\ \hline 
         \textbf{Model Prediction Time ($T_\text{predict}$) / $\text{s}$}&  392.8 &\textbf{17.3}&  $65.1$&2,693.8\\ \hline 
         \textbf{Mean Squared Error (MSE)}&  $4.2 \times 10^{-7}$ &0.931&  $(8.8 \pm 0.2) \times 10^{-6}$&$\mathbf{3.4\times 10^{-7}}$\\ \hline 
         \textbf{Mean Absolute Error (MAE)}&  0.00047 &0.908&  $(2.25 \pm 0.03) \times 10^{-3}$&$\mathbf{0.00036}$\\ \hline
 \textbf{Mean Absolute Percentage Error (MAPE) / \%}& 0.050 &90.1& $0.242 \pm 0.003$&$\mathbf{0.040}$\\\hline
    \end{tabular}
    \caption{Performance metrics of the benchmark CNN, DNN, the grid-optimised ELM, and the best ensemble predictor trained as a surrogate model for Experiment 2. Errors in the MSE, MAE and MAPE for the best ELM have been calculated as the standard deviation across different random initialisations of the ELM weights, corresponding to the top plot in Figure \ref{fig:expt-2-grid-search-2}. The benchmark, best ELM and best ensemble were all trained using the full train set of 6,400 samples. The best ELM used a hidden layer with 2,500 neurons, a Softplus activation function, and sampled the hidden weights from a uniform distribution. The best ensemble comprised 50 individual predictors, each corresponding to the hyperparameters of the best ELM, and differing in terms of random seed used to initialise the hidden layer weights. For each metric, the best value achieved across the benchmark CNN, DNN, the best ELM and the best ELM ensemble is highlighted in bold. The batch size for predictions using the CNN, DNN and ELM was 64.}
    \label{tab:results-summary-experiment-2}
\end{table*}

\section{Discussion}\label{sec:discussion}


A grid search across hyperparameters was conducted for both Experiment 1 and Experiment 2 to identify the best-performing ELM in each case.

In Experiment 1, the grid search was performed over all activation functions and hidden layer sizes ranging from 150 to 1,000 neurons, and the number of training samples used was 10,000 (0.28\% of the full training set used for the benchmark BIRNN). The results showed that all combinations of hyperparameters achieved lower MSE on the validation set compared to the benchmark model (Figure \ref{fig:expt-1-grid-search}).

For Experiment 2, the grid search was extended to include hidden layer sizes between 100 and 5,000, with hidden weight initializations sampled from both normal and uniform distributions (as shown in Figure \ref{fig:expt-2-grid-search}), and the full training set of 6,400 samples was used to train each model. However, none of the candidate ELMs achieved a lower test set MSE than the benchmark model. Consequently, the best-performing ELM from the grid search (with hyperparameters of 2,500 hidden neurons; a Softplus activation function, and weights sampled from a uniform distribution) was selected as the base predictor for an ensemble of 50 models, where each ensemble member differed by the random seed used to initialize the weight matrix. This ensemble model achieved an MSE of $3.4 \times 10^{-7}$ on the test set, outperforming the benchmark model by approximately 20\%.





\subsection{Comparison to Dense Neural Network}

In Figures \ref{fig:expt-1-ffnn} and \ref{fig:expt-2-ffnn}, the DNN shows higher prediction errors relative to the BIRNN in Experiment 1, and the CNN in Experiment 2, respectively.

In Experiment 1, where the task involves predicting vertical profiles of radiative flux from sequential low-dimensional inputs, the DNN exhibits wider interquartile ranges than the BIRNN and ELM, and less stable median errors than the ELM. This is especially evident in the 45–65 km region, where the presence of a cloud deck introduces added complexity to the radiative transfer modelling problem, and thus to the learning problem. The increased model error in this region is common across all models, but the DNN shows greater variability than the BIRNN and ELM. Below 45 km, all models exhibit reduced error, likely due to the decreased magnitude of flux at lower altitudes.

While the DNN architecture used in this study was designed with increased capacity (layer sizes of 512, 256, and 128 neurons) compared to the 128–128–128 architecture explored by \cite{ukkonen_exploring_2022}, this increase in parameter count did not yield notable performance gains relative to the BIRNN (as compared to the relative performance found by \cite{ukkonen_exploring_2022}). This suggests that model performance in this setting may be more influenced by architectural suitability—such as recurrent mechanisms for sequential data—than by depth or layer width alone, when considering the models which use gradient-based optimisation. Nonetheless, the DNN provides a useful baseline: it is architecturally simpler and less computationally intensive than recurrent networks, and its consistent performance across altitude highlights its utility as a general-purpose model.

In Experiment 2, we see in Figure \ref{fig:expt-2-ffnn} that the median prediction error of the DNN is much larger than that of all other models, and the distribution of the prediction errors is also much wider. We see a trend in the DNN prediction errors where certain parts of the lightcurve, such as between $\sim 0.6 -1.3\text{s}$ and $\sim 2.0-2.7\text{s}$ display larger median prediction errors compared to the rest of the lightcurve. The median prediction error of the DNN varies with time in a pattern that closely follows the median test lightcurve flux: this suggests that it may have not fully captured the localised relationships between stellar surface features and specific points in the lightcurve, and instead applies a more global mapping. In contrast, the ELM and CNN produce relatively constant error across the lightcurve, implying that they may more effectively leverage local features in the surface maps -- the CNN capturing these explicitly by architectural design, and the ELM in particular appearing to implicitly capture key spatial dependencies.

\subsection{Hyperparameter Choice}

Across the hyperparameter grid searches for Experiment 1 and Experiment 2 (Figure \ref{fig:expt-1-grid-search} and Figure \ref{fig:expt-2-grid-search} for Experiment 1 and Experiment 2 respectively) we see that there is only a small difference in model performance between the two different distributions from which hidden weights were sampled. We do however see that there is some level of consistency within each experiment, that the uniform distribution produces models with generally better performance than the normal distribution in Experiment 1, and that the normal distribution produces models with generally better performance in Experiment 2. 

We see that there is also only a small difference in model performance depending upon which activation function is used, though across both experiments we see that the Rectifier and Sigmoid activation functions both seem to produce models of slightly worse performance compared to most other activation functions. Referring to Figure \ref{fig:activation-functions} which displays the form of the activation functions, we suggest that the lower performance of the Rectifier is due to this activation function effectively deactivating around 50\% of the nodes, which the other activation functions do not do. Regarding the Sigmoid function, we note that this function is typically not used in regression tasks, such as this, which renders its lower performance unsurprising. The function has a lower sensitivity to small variations in the inputs for larger inputs, a feature which is not conducive to reliable performance in a regression task.


The number of neurons in the hidden layer seems to have a strong effect on model performance with a similar trend seen in both Experiment 1 and Experiment 2: the larger the hidden layer, the better the model performance. In Experiment 2, this trend occurs until the number of neurons in the hidden layer is 2,500, with errors increasing for models with 5,000 neurons in the hidden layer, as shown in Figure \ref{fig:expt-2-grid-search}. This can be seen for all activation functions except for the Rectifier function, likely for the aforementioned reason that the Rectifier function effectively `turns off' approximately half of the neurons in the ELMs. One might therefore expect that ELMs using the Rectifier activation function may still hit this point of minimum validation error but for a higher number of hidden layer neurons. The reason for ELMs with 2,500 neurons in the hidden layer outperforming those with 5,000 neurons in the hidden layer can likely be explained by the increased tendency to overfit the training dataset when the number of free model parameters approaches the number of samples in the training set, which is 6,400 for Experiment 2 -- this can be understood as the network `memorising' the training dataset instead of learning generalisable patterns within these examples. More formally, \cite{rudi_generalization_2017} investigate the generalisation properties of models learning with random features and via a convex objective, like the ELMs in this study, and find that for a training set of $n$ samples, \textit{"$O(1/\sqrt{n})$ learning bounds can be achieved with only $O(\sqrt{n}\log{n})$ random features"}, meaning that near-optimal generalisation rates can be achieved using a number of random features much smaller than the number of training samples. This gives some intuition as to why the ELM in Experiment 2 with 5,000 neurons does not reap significant performance gains over the ELM with 2,500 neurons.

\subsection{Random Seed for Weights Initialisation}

Figure \ref{fig:expt-1-grid-search-2} and Figure \ref{fig:expt-2-grid-search-2} display the model performance as a function of random seed for the hyperparameter-optimised ELM of Experiment 1 and Experiment 2 respectively. For Experiment 1, the ratio of the standard deviation of the model performances $\sigma$ to the mean performance $\bar{x}$ is 0.59\%, and for Experiment 2 we have $\sigma / \bar{x} = 2.68\%$. We see that for Experiment 1, the choice of random seed has a much lower effect on model performance compared to for Experiment 2: this potentially could be attributed to the larger number of parameters of the best ELM of Experiment 2 compared to Experiment 1, or it is perhaps related to the amount of variance within the data being modelled. For Experiment 1, the choice of random seed does not lead to any significant change in performance, though based on the result for Experiment 2 it may be advisable to train models across multiple random weights initialisations to enable potential increases in performance due to this variability.

\subsection{The Interplay of Data Dimensionality, Training Set Size, and Data Structure in ELM Performance}

Our experiments provide insights into how input dimensionality ($d$), training set size ($n_\text{train}$), and data structure influence ELM performance as surrogate models. Due to computational constraints in generating high-dimensional simulations, our experiments necessarily confound these variables, limiting causal inferences. However, the contrasting results offer valuable practical guidance for ELM application.

In Experiment 1, with sequential data and favorable dimensionality ratio ($d \ll n_\text{train}$), all tested ELMs substantially outperformed the benchmark RNN model. Conversely, in Experiment 2, with image data and unfavorable dimensionality ratio ($n_\text{train} \ll d$), individual ELMs failed to match the benchmark CNN's performance.

These performance differences likely stem from both dimensionality considerations and data structure. \cite{candes_near-optimal_2006} demonstrate that signal recovery from random projections depends critically on data "compressibility." The spatial dependencies in 2D image data become disrupted when flattened for ELM input, requiring more random projections (hidden neurons) to capture effectively compared to 1D sequential data. CNNs preserve these spatial relationships through their architectural design, explaining their advantage in Experiment 2.

Notably, our ELM ensemble in Experiment 2 outperformed the benchmark CNN, suggesting ensemble techniques can mitigate challenges associated with high-dimensional structured data. By combining multiple random projection spaces, the ensemble likely captures complementary aspects of the data structure, reducing variance and improving generalisation.

These findings suggest practical considerations for surrogate model selection. Beyond the $d/n_\text{train}$ ratio, practitioners should evaluate whether the inherent data structure is preserved in their modeling approach. ELMs appear particularly effective when random projections can adequately capture relevant patterns ($d \ll n_\text{train}$), while ensemble methods may extend their utility to more complex scenarios. Future work could systematically investigate how geometric data structure affects optimal ELM architecture, particularly hidden layer size, and explore methods for preserving structural information within random feature approaches.

\subsection{Sample Efficiency}

A crucial consideration in surrogate modeling for computationally expensive simulations is sample efficiency—how much training data is required to achieve satisfactory performance. Our experiments reveal a nuanced picture of ELM efficiency across different data scenarios.

In Experiment 1, the benchmark BIRNN required 3,584,000 training samples to achieve a validation MSE of $1.28\times10^{-4}$ (see Section \ref{sec:birnn-sample-efficiency} for training set size justification). In contrast, Figure \ref{fig:expt-1-sample-efficiency} shows that an ELM trained with 5,000 samples (0.14\% of the data used by the BIRNN) achieved approximately a six-fold reduction in validation MSE. Further performance improvements were observed as training set size increased across the range of training set sizes that were tested (though at some point these performance gains would likely plateau).

In Experiment 2, however, individual ELMs did not demonstrate the same advantage. None of the single ELMs matched the CNN benchmark regardless of training set size, highlighting important limitations to ELM performance in high-dimensional image data compared to low-dimensional sequential data. Only when combined in an ensemble did ELMs show efficiency benefits. As shown in Figure \ref{fig:experiment-2-sample-efficiency}, an ensemble of 50 ELMs trained on 5,000 examples (78\% of the full training set) marginally outperformed the benchmark CNN that utilised all 6,400 available training samples. This more modest efficiency gain required the additional computational cost of training multiple models, though still at a much reduced cost of $16.4\times$ less CPU compute time compared to training a CNN (see Table \ref{tab:results-summary-experiment-2}).

These contrasting results suggest that ELM sample efficiency varies significantly with data characteristics. In lower-dimensional sequential data (Experiment 1), the direct analytical solution method of ELMs likely avoids the overfitting and optimisation challenges faced by gradient-based learning, enabling superior performance with drastically fewer samples. For complex high-dimensional image data (Experiment 2), this advantage diminishes, requiring ensemble techniques to compensate.

For exoplanet science applications, where generating simulation data may require substantial computational resources, these findings suggest a context-dependent approach. In scenarios resembling Experiment 1, ELMs offer transformative efficiency gains, potentially reducing required simulations by over 99\%. In scenarios closer to Experiment 2, practitioners should carefully weigh the trade-offs between using more sophisticated architectures like CNNs versus implementing ELM ensembles, considering both data generation costs and accuracy requirements.

The differential sample efficiency across experiments aligns with our earlier discussion on dimensionality and data structure, reinforcing the importance of considering these factors when selecting appropriate surrogate modeling approaches.

\subsection{Training Time}

For both Experiment 1 and Experiment 2, the best candidate ELM-based models reap a significant reduction in training time compared to the deep learning benchmark models. 

For Experiment 1, the candidate ELM performing with the lowest validation set MSE has a training time $T_\text{train} = 0.45 \, \text{s}$ on one CPU core, compared to the benchmark BIRNN for which $T_\text{train} = 146,716.66 \, \text{s} $ (approximately 41 hours) on one CPU core. Training the BIRNN on one NVIDIA A100 GPU core reduces the training time to approximately 12 hours (see Table \ref{tab:benchmark-gpu-training-times}). In either case, we see a significant reduction in the training time and potentially computational resources required to train the ELM surrogate as compared to the BIRNN benchmark, with the ELM achieving a $>300,000\times$ reduction in training time as compared to the BIRNN when using one CPU core. 

For Experiment 2, the best-performing ELM Ensemble took $9,216.16 \, \text{s}$ to train on one CPU core, which vastly undercuts the training time of the benchmark CNN, which was $151,413.6 \, \text{s}$ on one CPU core. This represents a $16.4\times$ reduction in training time when using the ensemble of ELMs compared to using the CNN. The ELM Ensemble training time on one CPU is also $5.7\times$ faster than the CNN training time on one GPU, shown in Table \ref{tab:benchmark-gpu-training-times}.

Comparing the ELM to the DNN, we see a $7800\times$ reduction in training time in Experiment 1, and a much smaller though still substantial $35\times$ reduction in training time in Experiment 2.

In both experiments, we observe a significant reduction in the training time when using ELMs and ELM Ensembles as compared to deep learning alternatives. It is important to note that different batch sizes were used across models due to practical constraints such as input dimensionality. For example, the BIRNN and DNN in Experiment 1 used batch sizes of 512 and 1024, respectively, while the CNN and DNN in Experiment 2 used batch sizes of 64 and 8. As such, reported training times should be interpreted with this context in mind, and not as strict like-for-like comparisons. Nonetheless, even accounting for these considerations, the scale of difference in training time between ELM-based models and deep learning benchmarks remains substantial.

\subsection{Prediction Time}

In both Experiment 1 and Experiment 2, all ELMs across the grid searches improve upon the prediction time $T_\text{predict}$ of the deep learning counterparts (for ELM prediction times, see Figure \ref{fig:expt-1-grid-search-2} for Experiment 1; Figure \ref{fig:expt-2-grid-search-2} for Experiment 2). For Experiment 1, the best ELM (that with the lowest MSE on the validation set) has $T_\text{predict} = 1.14 \, \text{s}$, compared to for the BIRNN benchmark which has $T_\text{predict} = 41.78 \, \text{s}$ (both measured on one CPU core for 50,000 samples), equating to a $36\times$ reduction in prediction time. 

For Experiment 2, though the individual ELMs offer a reduction in prediction time as compared to the CNN benchmark, the best ELM Ensemble results in an increased prediction time of $T_\text{predict} = 2,693.76 \, \text{s}$ compared to the CNN's $T_\text{predict} = 392.8 \, \text{s}$; a significant $6.9\times$ increase in prediction time. The ensemble prediction time was measured on one CPU core, with the total prediction time potentially being reduced by using an increased number of parallel processes.

Comparing the ELM to the prediction time of the DNN, we see marginal speed-up in Experiment 1, and increase in prediction time in Experiment 2 by about $4\times$. In the latter case, this slow-down is justified as the DNN tested does not achieve adequate test performance.

\subsection{Ensemble Performance}

\subsubsection{Performance of ELM Ensemble vs. Individual ELMs}


In Experiment 2, we synthesize both individual ELM and Ensemble ELM surrogate models for the image-processing task. Our results demonstrate superior performance of the Ensemble ELM compared to individual ELM models. This finding aligns with existing empirical research on ELM ensembles \citep{lan_ensemble_2009, liu_ensemble_2010, sun_os-elm_2011}, which consistently shows that Ensemble ELMs deliver greater stability and accuracy than standalone ELM models. Intuitively, by averaging predictions across multiple ELM models, the ensemble mitigates the inherent randomness introduced by the stochastic initialisation of hidden layer weights in individual ELMs, leading to more robust predictions.


We did not investigate Ensemble ELMs in Experiment 1 as individual ELMs performed with far superior accuracy compared to the benchmark BIRNN, and ensembling introduces increased cost to training and prediction time. Based upon findings of Ensemble ELM performance compared to individual ELM predictors, we hypothesise that Ensemble ELMs would offer prediction accuracy comparable to or improved compared to the standalone ELM predictors investigated in Experiment 1, though at a cost of increased training and prediction times, which in many cases can be offset with the use of parallelization.

\subsubsection{Performance of ELM Ensemble vs. CNN Ensemble}

To better understand the effect of ensembling on model performance in Experiment 2, we conducted a focused comparison between CNN and ELM ensembles with a maximum of 15 predictors. This constraint was necessary to make CNN ensembling computationally feasible and ensures a fair, matched comparison between the two methods. The results in Figure \ref{fig:ensemble-performance} demonstrate that both ensemble types benefit from increasing ensemble size, with test error consistently decreasing as more predictors are added. Initially, the ELM ensemble shows a slightly steeper improvement with the first few predictors; however, beyond approximately 8 predictors, both methods exhibit a similar rate of performance gain. Despite this convergence in trend, the CNN ensemble maintains a consistent advantage in absolute test accuracy across all ensemble sizes. This suggests that while both models benefit comparably from ensembling beyond a certain size, CNNs offer stronger individual performance, which accumulates to a greater overall effect when ensembled.

However, these performance differences come with significant practical trade-offs. CNNs are computationally expensive to train, with each model requiring $\sim700 \times$ more CPU training time $\sim6 \times$ more CPU inference time than an ELM. For prediction, both ensemble types incur increased inference cost with ensemble size, but the cost per model remains significantly lower for ELMs. As such, while CNN ensembles offer stronger performance, ELM ensembles provide a highly efficient alternative when computational resources or latency constraints are a concern. This highlights a key trade-off between accuracy and efficiency that should be considered when selecting models for deployment.

\subsection{Limitations and Scope of Study}


This work evaluates the performance of ELM surrogates as compared to state-of-the-art deep learning algorithms for exoplanet-related simulation tasks. Here, we select two tasks which deal with different types of data structures: sequentially-structured data for Experiment 1, and image-structured data for Experiment 2. The ELM does not have the geometric structure of the data explicitly represented in the inputs or in the model architecture, and must instead process the geometric structure of the data implicitly. We note that owing to the context of producing surrogate models for exoplanet simulations, the training datasets are not exhaustively large, as data generation cost is a fundamental consideration when synthesising surrogate models for these simulations.

We find that our results are consistent with findings from \cite{markowska-kaczmar_extreme_2021}, who investigate whether randomisation-based learning can be competitive with neural networks trained using backpropagation, and who conclude that ELMs perform better than networks trained via backpropagation for relatively small datasets, but are not competitive in image processing tasks.  We acknowledge that the performance advantage of ELMs over BIRNNs may diminish or reverse with larger datasets, though our study deliberately operates within practical data generation constraints typical of simulation surrogate modeling. Our conclusions are limited to these smaller training regimes.

We note that the image processing task we use in Experiment 2 is a relatively simple task, as the images being processed are of binary intensities (as displayed in Figure \ref{fig:experiment-2-input-target}). We hypothesise that the performance gap between ELMs and CNNs would likely widen with increasing image complexity, though this remains to be explored in future research.

Though we show that ELM surrogate models are not competitive against CNNs in the specific image-processing task of this study, ELM Ensembles achieve comparable performance with significantly reduced training time, albeit at the cost of increased prediction time. This presents an opportunity for surrogate modeling of image-processing tasks in time-sensitive contexts where training efficiency outweighs prediction speed. We also note that because of the fact that our image-processing task in Experiment 2 is fairly simple, more complex image-processing tasks may require an ensemble with more free parameters in order to achieve comparable performance to a CNN.


There is a key discrepancy between all gradient-optimised models in this work: the batch size used during training varies highly between the CNN, BIRNN and DNN. This limits the comparability of training times quoted within Table \ref{tab:results-summary-experiment-1} and Table \ref{tab:results-summary-experiment-2}. Batch sizes were selected based on input dimensionality and available memory for each model. As such, differences in training time reflect both inherent model efficiency and these practical limitations. Reported training times should be interpreted with this in mind, rather than as direct comparisons under identical computational conditions. Furthermore, in both experiments, the chosen batch sizes may have limited GPU efficiency and consequently the GPU training times quoted in Table \ref{tab:benchmark-gpu-training-times} may not represent the optimal GPU training times. For the CNN in Experiment 2, a small batch size of 8 was likely suboptimal given the large input size and model complexity; larger batch sizes (e.g., 32–64) may have improved training speed. Similarly, the BIRNN in Experiment 1 used a batch size of 512, which may not have fully utilised GPU resources. These choices of batch size could explain the modest speedups observed on GPU. Future work will explore more efficient batch size configurations.

To strengthen conclusions about the applicability of certain emulation strategies to simulations in exoplanet science, future work could involve testing these methods on widely used benchmark datasets. Examples include the dataset from \cite{marquez-neila_supervised_2018}, which has also been used in \cite{cobb_ensemble_2019} and \cite{matchev_analytical_2022}, as well as the dataset from \cite{himes_accurate_2022}. This represents a promising direction for continued investigation into emulation strategies for exoplanet modelling.





\section{Conclusions}\label{sec:conclusion}

This comparative study evaluates Extreme Learning Machines (ELMs) against established deep learning benchmarks for use as surrogate models in exoplanet simulations. Our findings reveal ELMs' context-dependent strengths and limitations across different simulation types, with performance patterns closely tied to data dimensionality, structure, and training set size.

In sequentially-structured data processing (Experiment 1), where input dimensionality was substantially lower than training set size, ELMs demonstrated superior performance compared to BIRNN benchmarks. Our optimally-performing ELM achieved better accuracy while requiring only 0.3\% of the training examples needed by the benchmark model. Additionally, ELMs significantly reduced both training time (by $100,000\times$) and inference latency in these scenarios.

For image-based data processing (Experiment 2), characterised by high dimensionality and more complex spatial relationships than that of Experiment 1, individual ELMs underperformed relative to the benchmark CNN. However, when implemented as ensembles, ELMs matched CNN accuracy while training approximately 16 times faster. This performance advantage came at the cost of inference speeds approximately 7 times slower than their CNN counterparts, though there is potential to accelerate both ensemble training and prediction by parallelising over multiple CPU cores.

Our results lead to several important insights regarding the application of ELMs in exoplanet simulations:
\begin{enumerate}
    \item ELMs show particular promise for low-dimensional sequential data mapping problems, where their random projection approach can rapidly and successfully emulate simulations with remarkable sample efficiency. This suggests that using ELMs as surrogates for such simulations could substantially reduce the computational expense of data generation compared to conventional deep learning approaches.
    \item While ELM ensembles can effectively match CNN performance for image processing simulation such as that of Experiment 2, their slower inference times necessitate careful consideration for time-sensitive applications, suggesting a trade-off between training efficiency and operational speed.
    \item The minimal computational resources required for training ELMs offers several practical advantages: it enables the integration of surrogate training directly into simulation runtime; broadens access to surrogate modeling for research groups without high-performance computing facilities; and facilitates more rapid surrogate model prototyping.
\end{enumerate}

These findings highlight the potential of ELMs as valuable tools in scientific modeling, particularly for specific data structures where their strengths in sample efficiency and training speed can be leveraged effectively. Our results suggest that sequential data processing tasks in atmospheric modeling and other domains with similar data characteristics stand to benefit most substantially from ELM implementation. Future work should consider how geometric data structure affects optimal ELM architecture and explore methods for preserving structural information within random feature approaches to extend ELM applicability to more complex simulation domains.


\section*{Acknowledgments}

The authors acknowledge the use of the High Performance Computing facilities of University College London to carry out this work, specifically the Hypatia cluster. This work used computing equipment funded by the Research Capital Investment Fund (RCIF) provided by UKRI, and partially funded by the UCL Cosmoparticle Initiative.
This research received funding from the European Research Council (ERC) under the European Union's Horizon 2020 research and innovation programme (grant agreement n$^\circ$ 758892/ExoAI), and from the Science and Technology Facilities Council (STFC; grant n$^\circ$ ST/W00254X/1 and grant n$^\circ$ ST/W50788X/1). JMM acknowledges support from the Horizon Europe Guarantee Fund, grant EP/Z00330X/1.

\section*{Data Availability}

The data used to train and evaluate the models in this work have been generated by the methods outlined in Section \ref{sec:experiments}.



\bibliographystyle{rasti}
\bibliography{references} 




\appendix

\section{Selection of Benchmark Models}\label{sec:selection-of-benchmark-models}

\subsection{Experiment 1: Bi-directional Recurrent Neural Network}

The benchmark deep learning model for the task of two-stream atmospheric radiative transfer has been chosen as a Bi-directional Recurrent Neural Network, or BIRNN. This is based upon work by \cite{ukkonen_exploring_2022}, which is a comparative study of various candidate deep learning architectures for use as a surrogate model for two-stream radiative transfer, tested on simulations of the Earth's atmosphere. This work demonstrated superior performance of a BIRNN compared to a Feedforward Neural Network for the task at hand, and informed the architecture used for emulation of the two-stream radiative transfer in the \texttt{OASIS} Global Circulation Model \citep{mendonca_new_2014} in our paper \cite{tahseen_enhancing_2024}, where our model achieved performance consistent with the model described in \cite{ukkonen_exploring_2022}. The model developed in \cite{tahseen_enhancing_2024} is used as the benchmark deep learning model in this study for comparison with the ELM surrogate models.

\subsection{Experiment 2: Convolutional Neural Network}\label{sec:selection-of-cnn}

This appendix details the rationale and methodology behind the development of our custom Convolutional Neural Network (CNN) architecture used as the benchmark model in Experiment 2.

The architecture follows an AlexNet-inspired design \citep{krizhevsky_imagenet_2012}, with two sequential convolutional layers followed by a max pooling layer. Each convolutional layer employs a kernel size of 3. During model development, we considered output filters ranging from 2 to 32 in powers of 2, ultimately selecting 4 filters to maintain model parsimony while preserving performance.

We also explored incorporating both batch normalization and attention mechanisms. However, empirical testing revealed superior performance with fewer parameters, leading us to adopt a more streamlined configuration. The CNN architecture was specifically selected for its inductive bias toward image-based data processing, which aligns with the spatial nature of our task.

Hyperparameter optimization was conducted using a grid search. Upon identifying a subset of models demonstrating competitive performance, we applied Occam's razor, selecting the simplest architecture from among the top performers. It should be noted that our implementation does not utilize padding during image down-sampling operations.

While we conducted comprehensive hyperparameter searches, we acknowledge that the optimal model may lie outside the explored parameter space. Nevertheless, we contend that the pursuit of a single, definitively optimal model is not essential to our research objectives. The selected architecture demonstrates very good performance on its training objectives and is adequately suited to the needs of the task.

\section{Activation Functions}\label{appendix:activation-functions}

In this section we detail the activation functions used in this work, which are as follows:

\begin{enumerate}
\item \textbf{Sigmoid}:
\[
\text{Sigmoid function: } \sigma(x) = \frac{1}{1 + e^{-x}}
\]

\item \textbf{Softsign}
\[
\text{Softsign function: } f(x) = \frac{x}{1 + |x|}
\]

\item \textbf{Hyperbolic Tangent (Tanh)}
\[
\text{Tanh function: } \tanh(x) = \frac{e^x - e^{-x}}{e^x + e^{-x}}
\]

\item \textbf{LeCun Tanh}
\[
\text{LeCun Tanh function: } f(x) = 1.7159 \cdot \tanh\left(\frac{2}{3}x\right)
\]

\item \textbf{Softplus}
\[
\text{Softplus function: } f(x) = \ln(1 + e^x)
\]

\item \textbf{Rectifier}
\[
\text{Rectifier function: } f(x) = \max(0, x)
\]

\item \textbf{Gaussian}
\[
\text{Gaussian function: } f(x) = e^{-x^2}
\]
\end{enumerate}

\begin{figure}[!ht]
    \centering    \includegraphics[width=1\linewidth]{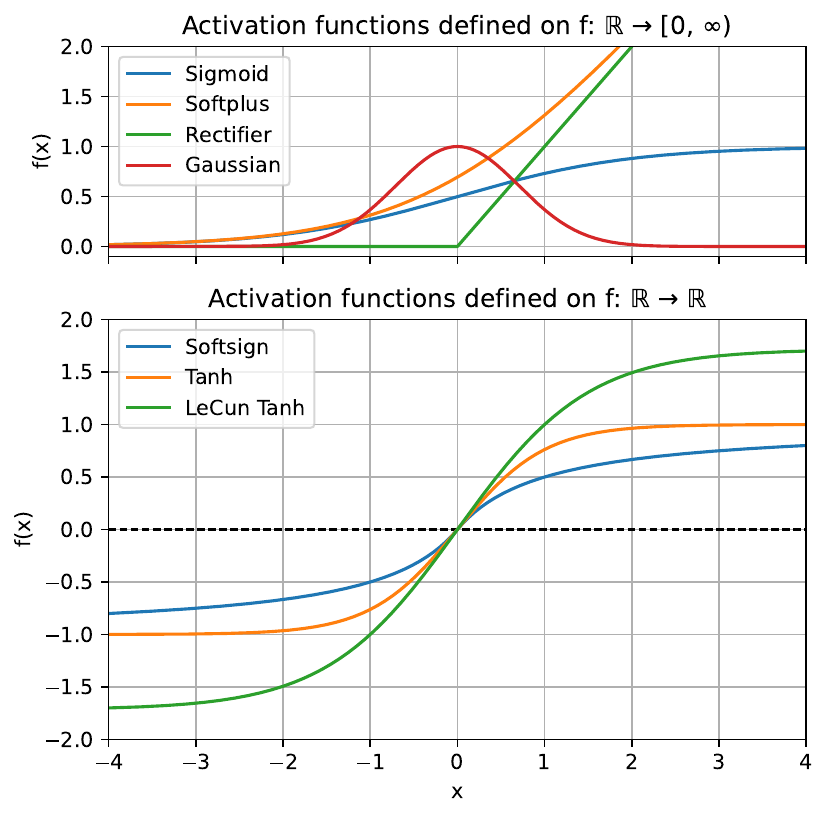}
    \caption{Activation functions used in the experiments.}
    \label{fig:activation-functions}
\end{figure}

These activation functions are plotted across a finite range in Figure \ref{fig:activation-functions}.

\section{Distributions}\label{appendix:distributions}

The hidden layer weights $\mathbf{W}$ defined in Equation \ref{eq:W} were randomly initialised using two different initialisation strategies:
\begin{enumerate}
    \item \textbf{Normal distribution: }
The weights were sampled independently from a normal distribution, such that $W_{ij} \sim \mathcal{N}(\mu, \sigma^2)$ where $\mu=0$ and $
\sigma^2=\sqrt{\frac{1}{m}}$, where $m$ is the dimensionality of the input vector to the ELM. Explicitly, we have for each parameter of $\mathbf{W}$: 
\begin{equation}
P(W_{ij}) = \frac{1}{\sqrt{2\pi\sigma^2}}e^{- \frac{(W_{ij}-\mu)^2}{2\sigma^2}}
\end{equation} 
    \item \textbf{Uniform distribution:}
The weights were sampled independently from a uniform distribution, such that  $W_{ij} \sim \mathcal{U}(-a,a)$ where $a=\sqrt{\frac{1}{d}}$. The probability distribution on each parameter of $\mathbf{W}$ is then:
    \begin{equation}
    P(W_{ij}) =
\begin{cases} 
\frac{1}{2a}, & \text{if } W_{ij} \in [a,-a] \\
0, & \text{otherwise}
\end{cases}
\end{equation}
\end{enumerate}
When initialising weights and biases, the bias term is absorbed into the weights matrix as an additional dimension, and these two quantities are initialised together using the same random seed. Effectively this means we redefine Equation \ref{eq:H} as $\mathbf{H} = g(\mathbf{X'W'}^{T})$ where $\mathbf{X'}=[\mathbf{X}, 1]$ and $\mathbf{W'}=[\mathbf{W}, \mathbf{b}]$ and then initialise $\mathbf{W}'$ as per the procedure above.



\section{Benchmark Model Training Times}\label{sec:benchmark-model-gpu-training-times}

In Section \ref{sec:results}, we compare surrogate model training times for models trained on a single CPU. This comparison is made in order to enable an apples-to-apples comparison of the training resource requirement for the ELM surrogates compared to the benchmark models. We acknowledge that deep learning surrogate models are likely to be trained on Graphic Processing Unit (GPU) resources, and we quote in Table \ref{tab:benchmark-gpu-training-times} the time duration of training the benchmark models on an NVIDIA A100 GPU.

\begin{table}[h!]
    \centering
    \begin{tabular}{ccc}
          Experiment&Model& Training Time / s\\
          1&BIRNN& 46,154.8 \\
          2&CNN& 53,259.4\\
    \end{tabular}
    \caption{Benchmark model training times on one NVIDIA A100 GPU.}
    \label{tab:benchmark-gpu-training-times}
\end{table}

\section{BIRNN Sample Efficiency}\label{sec:birnn-sample-efficiency}
\begin{figure}[!ht]
    \centering    \includegraphics[width=1\linewidth]{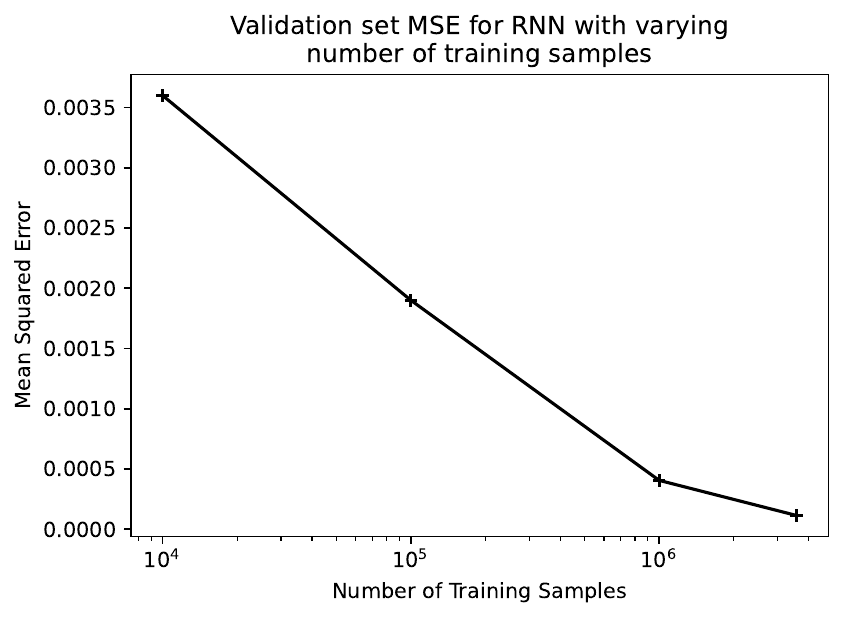}
    \caption{Mean squared error of BIRNN models on the validation set, whereby models are trained with varying number of training samples. We observe MSE on the validation set to decrease for increasing number of training samples.}
    \label{fig:rnn-sample-efficiency}
\end{figure}

Figure \ref{fig:rnn-sample-efficiency} plots the sample efficiency of the BIRNN benchmark, using MSE on the validation set to quantify model performance. We observe that validation set MSE decreases with increasing number of samples, with this trend not levelling off below 3,584,000 samples, as we see an appreciable decrease in the MSE when increasing training set size from 1,000,000 to 3,584,000 samples. This validates the use of a BIRNN with 3,584,000 training samples -- a large training set compared to those used for training the ELM -- as an appropriate benchmark with which to compare the training speed and sample efficiency of the ELM against. 

\onecolumn
\section{Results}
\subsection{Experiment 1}

\begin{figure*}[h!]
  \centering
  \includegraphics[width=.45\textwidth]{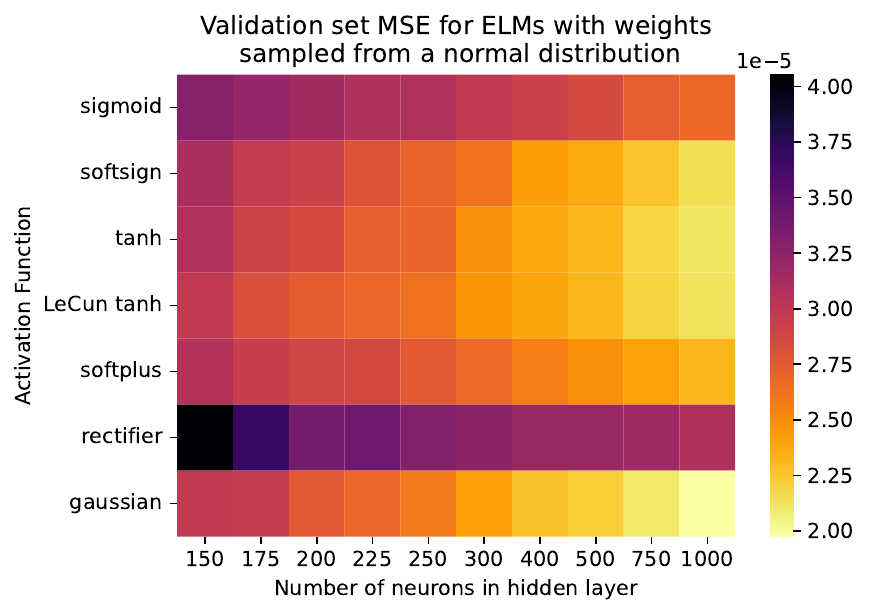}
  \hspace{.5cm}
  \includegraphics[width=.45\textwidth]{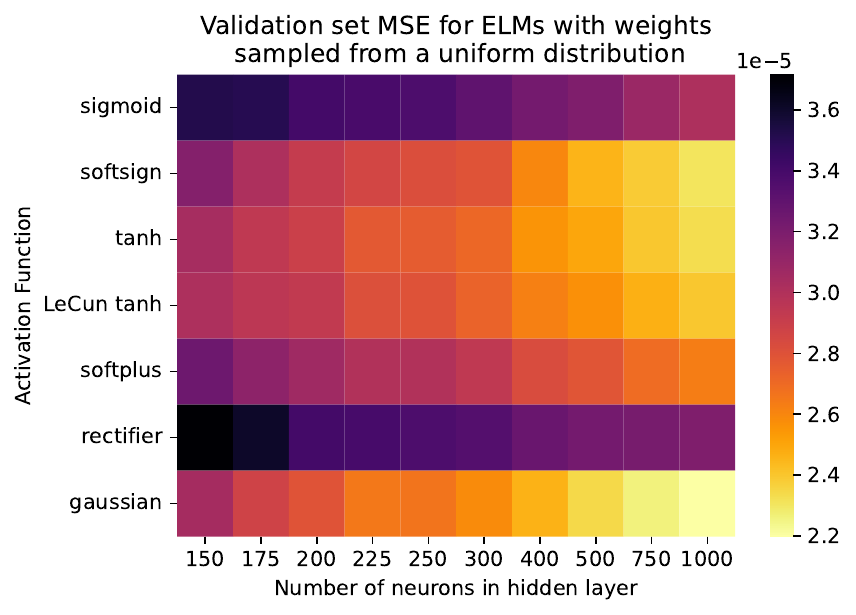}

  \vspace{.5cm}

  \includegraphics[width=.45\textwidth]{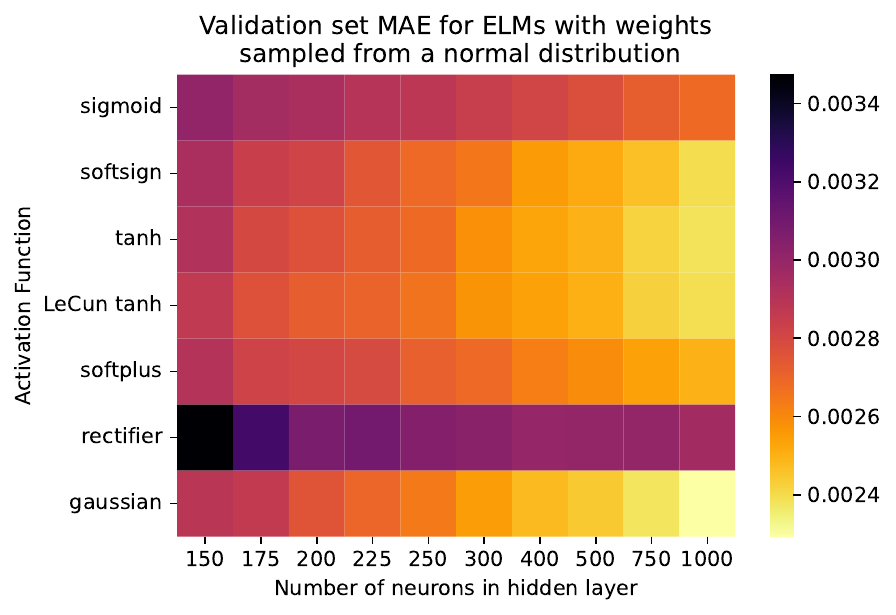}
  \hspace{.5cm}
  \includegraphics[width=.45\textwidth]{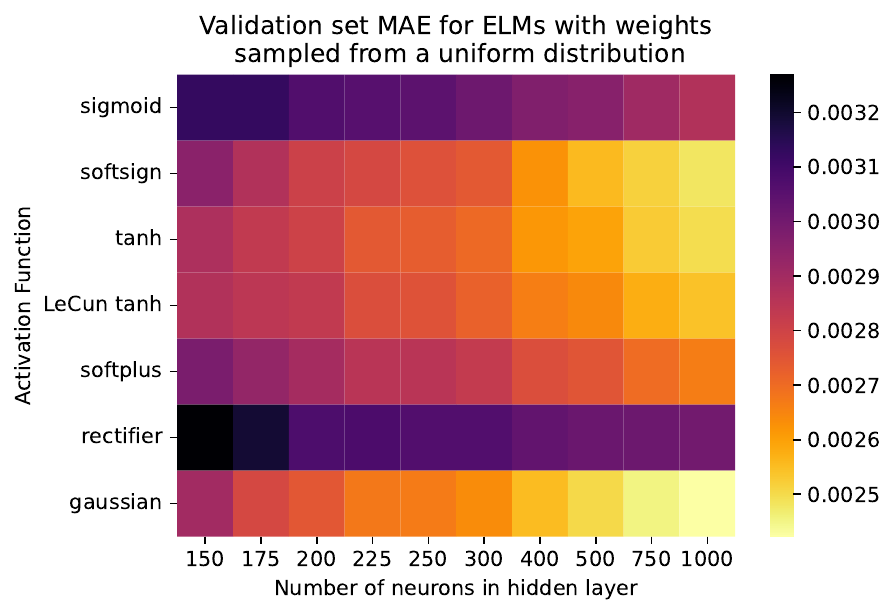}

  \vspace{.5cm}
  \includegraphics[width=.45\textwidth]{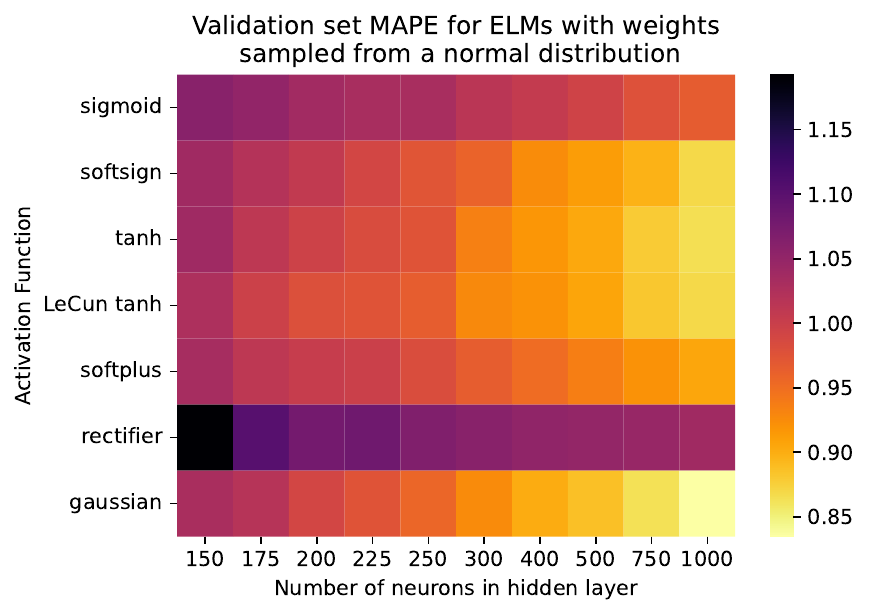}
  \hspace{.5cm}
  \includegraphics[width=.45\textwidth]{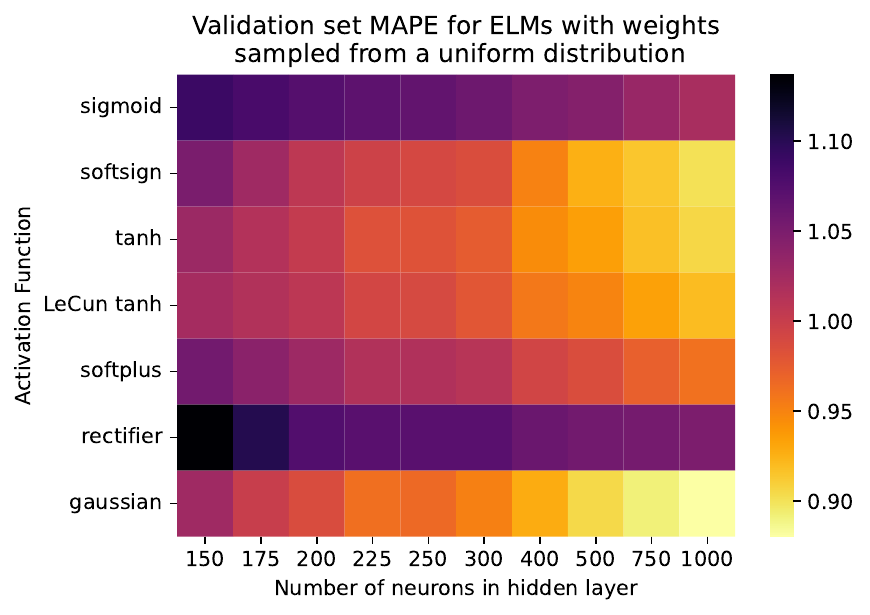}

  \caption{Mean squared error (top), mean absolute error (middle), and mean absolute percentage error (bottom) acorss ELMs trained with different activation functions and varying hidden layer size, for weights initialised from a normal distribution (left), and from a uniform distribution (right), for Experiment 2.}
  \label{fig:expt-1-grid-search}
\end{figure*}

\begin{figure*}[h!]
  \centering
  \includegraphics[width=.45\textwidth]{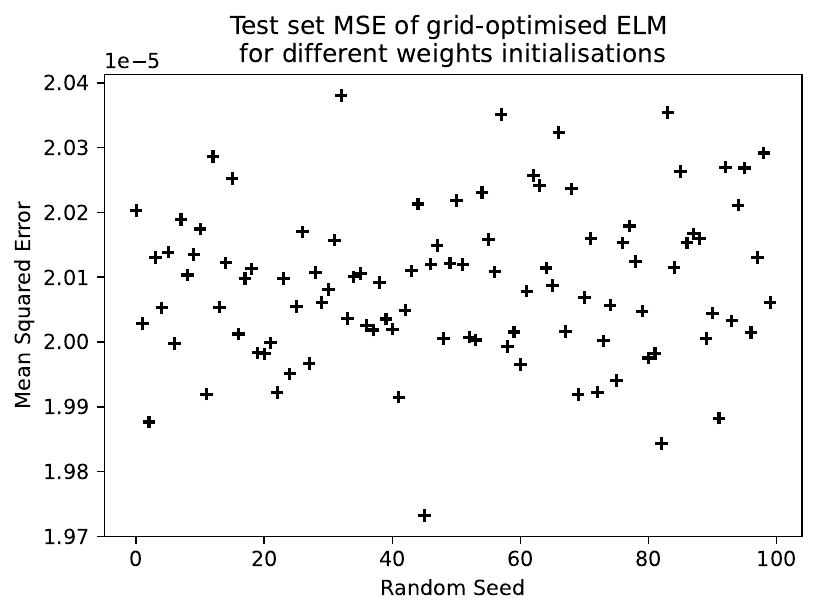}

  \vspace{.5cm}

  \includegraphics[width=.45\textwidth]{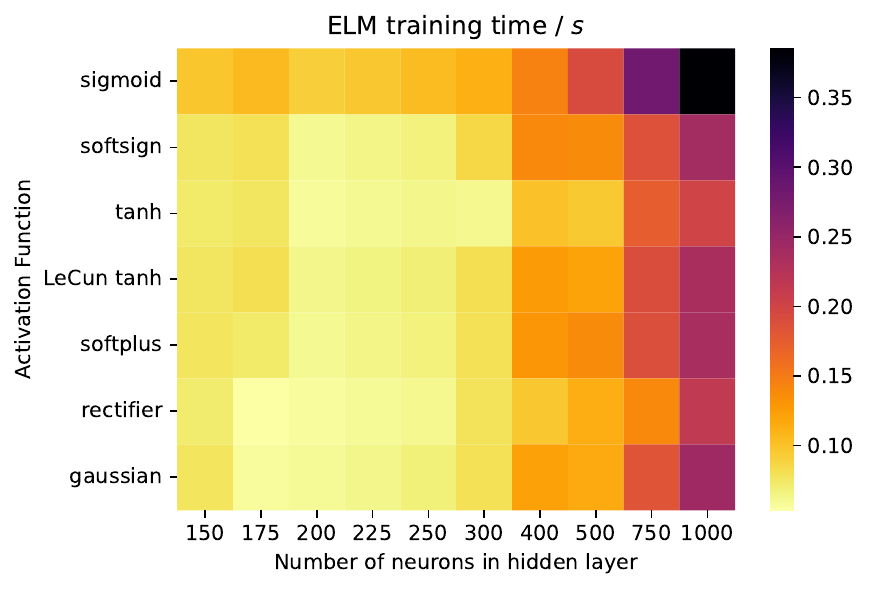}
  \hspace{.5cm}
  \includegraphics[width=.45\textwidth]{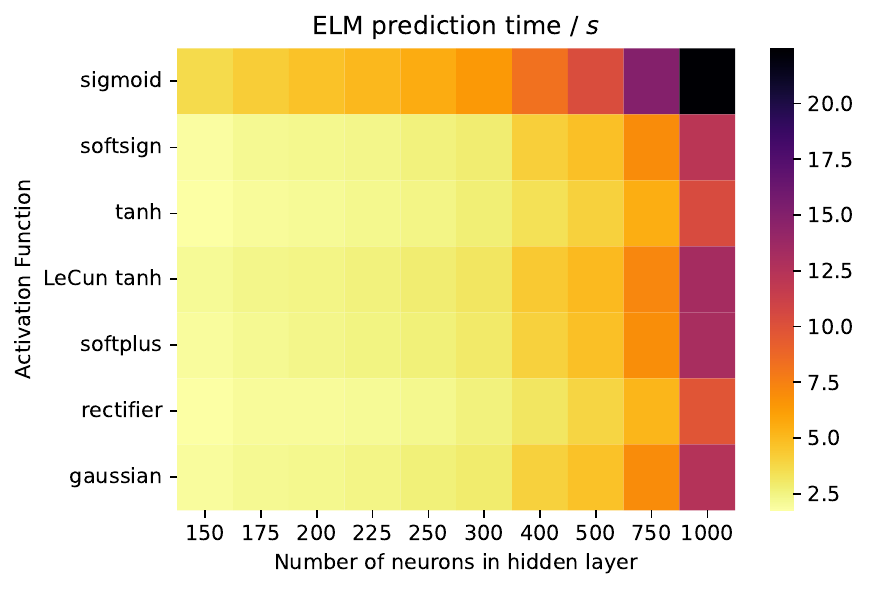}
  


  \caption{\textbf{Top:} Mean squared error on the validation set for an ELM with the best performing set of hyperparameters in the grid search (as described in Table \ref{tab:optimal-hyperparameters-table}) across varying random seeds used to initialise weights, for Experiment 1.
  \\
  \textbf{Bottom:} Training time (left) and prediction time (right) for ELMs with varying activation function and hidden layer size, for Experiment 1.
  }
  
    \label{fig:expt-1-grid-search-2}
\end{figure*}





\newpage
\onecolumn
\subsection{Experiment 2}

\begin{figure*}[h!]
  \centering
  \includegraphics[width=.45\textwidth]{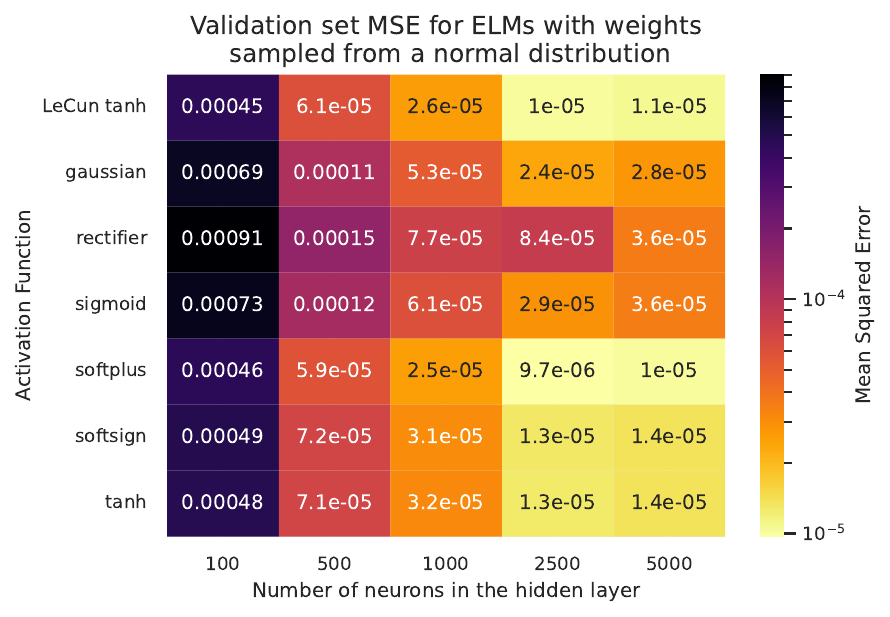}
  \hspace{.5cm}
  \includegraphics[width=.45\textwidth]{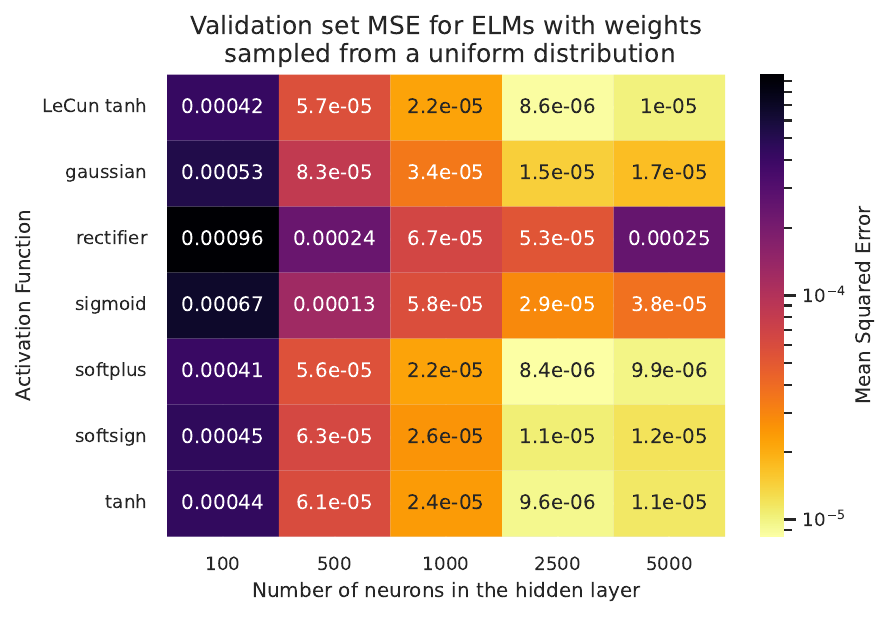}

  \vspace{.5cm}

  \includegraphics[width=.45\textwidth]{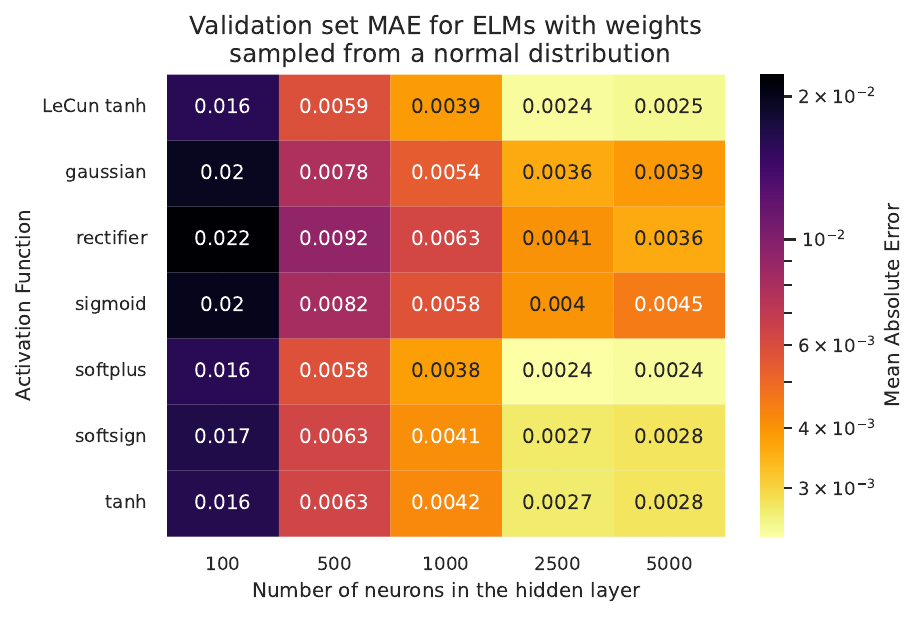}
  \hspace{.5cm}
  \includegraphics[width=.45\textwidth]{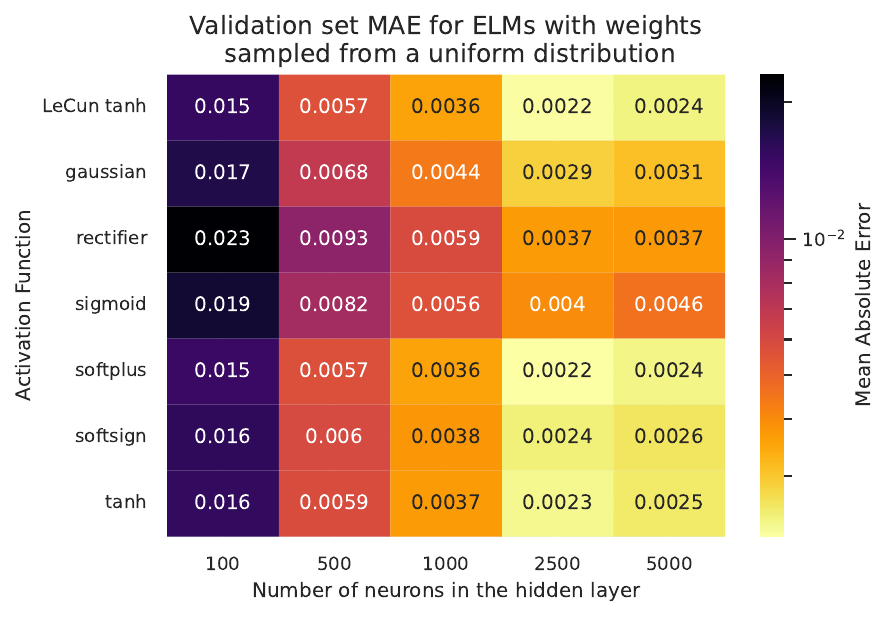}

  \vspace{.5cm}
  \includegraphics[width=.45\textwidth]{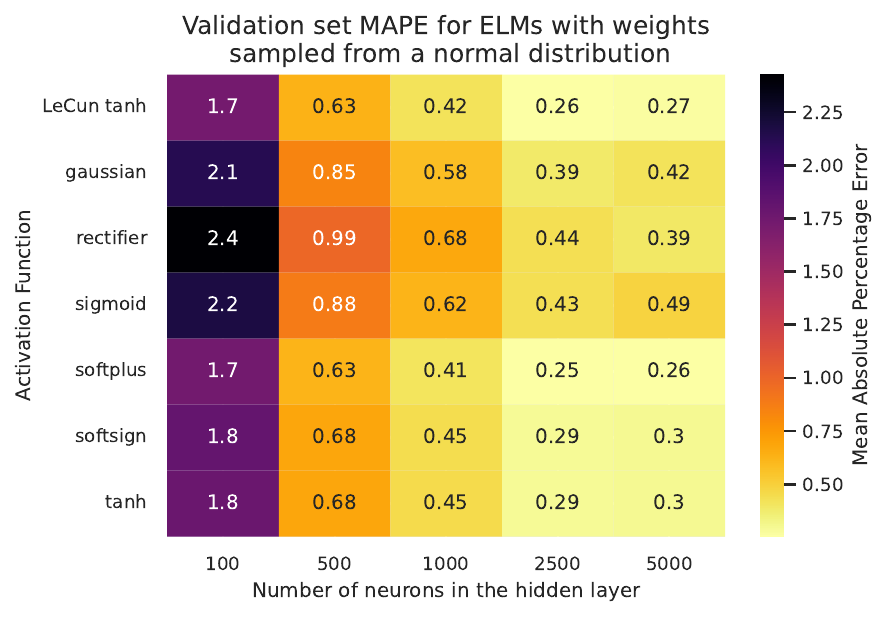}
  \hspace{.5cm}
  \includegraphics[width=.45\textwidth]{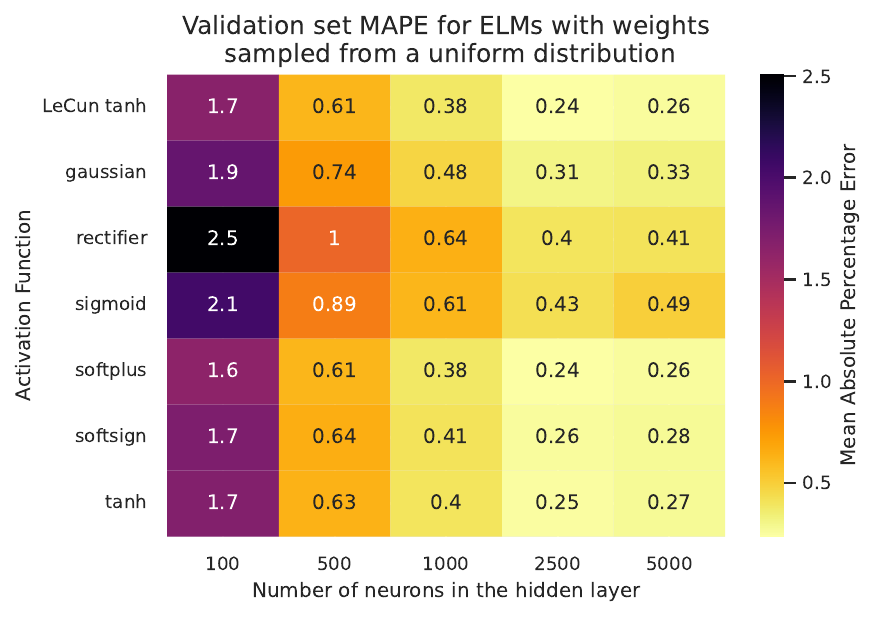}

  \caption{Mean squared error (top), mean absolute error (middle), and mean absolute percentage error (bottom) acorss ELMs trained with different activation functions and varying hidden layer size, for weights initialised from a normal distribution (left), and from a uniform distribution (right), for Experiment 2.}
  \label{fig:expt-2-grid-search}
\end{figure*}



\begin{figure*}[h!]
  \centering
  \includegraphics[width=.45\textwidth]{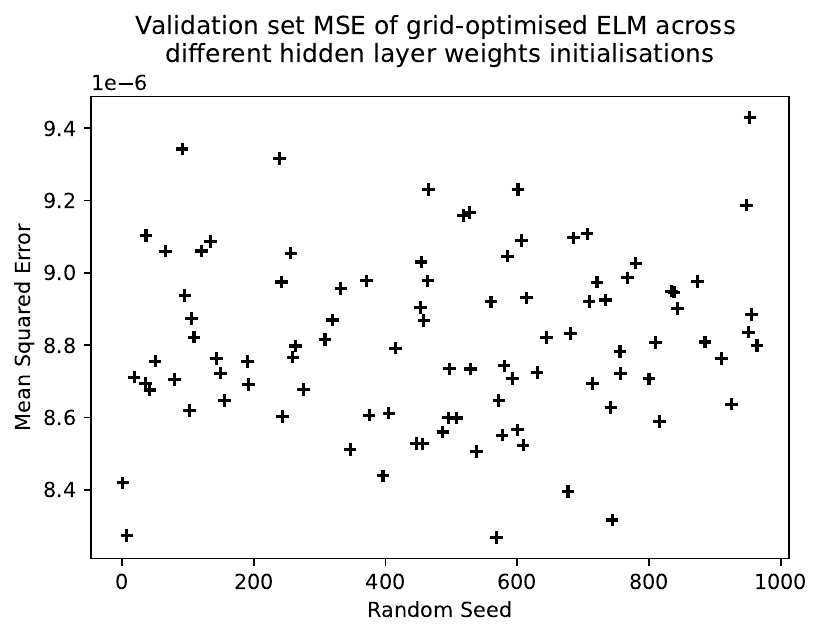}
  \hspace{.5cm}

  \vspace{.5cm}

  \includegraphics[width=.45\textwidth]{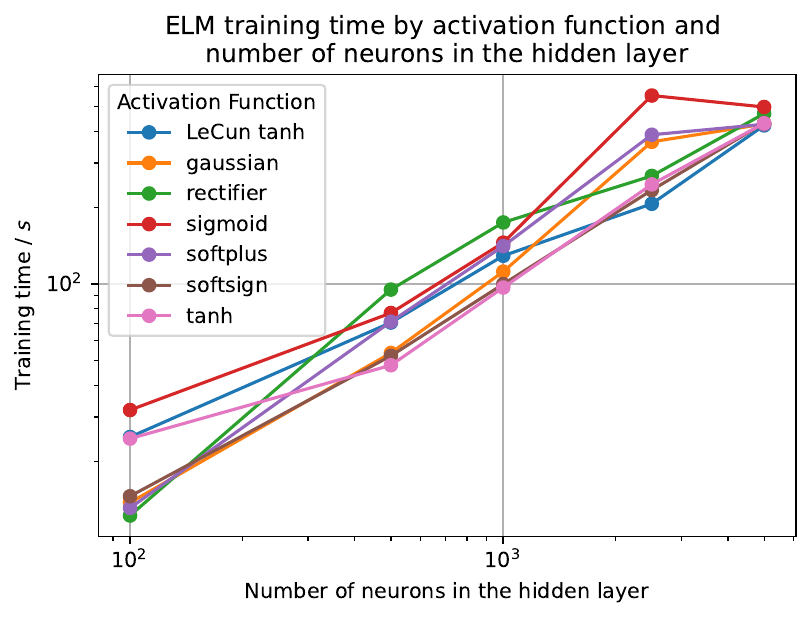}
  \hspace{.5cm}
  \includegraphics[width=.45\textwidth]{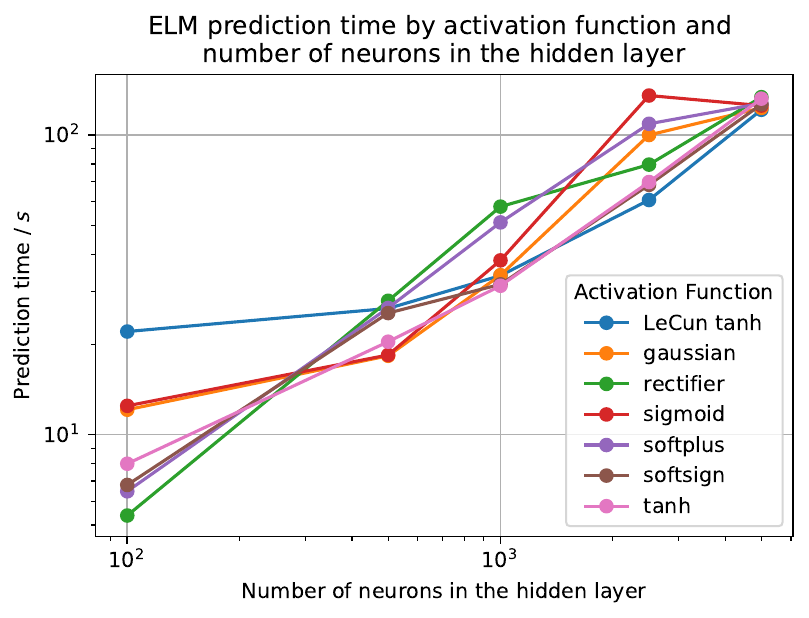}


  \caption{\textbf{Top:} Figure shows the mean squared error on the validation set for an ELM with the best performing set of hyperparameters in the grid search (as described in Table \ref{tab:optimal-hyperparameters-table}) across varying random seeds used to initialise weights, for Experiment 2.
  \\
  \textbf{Bottom:} Figure shows the training time (left) and prediction time (right) for ELMs with varying activation function and hidden layer size, for Experiment 2.
  }
  \label{fig:expt-2-grid-search-2}
\end{figure*}

\bsp	
\label{lastpage}
\end{document}